\documentstyle[11pt]{article}
\def\beq{\begin{equation}}
\def\eeq{\end{equation}}
\def\beqx{\begin{displaymath}}
\def\eeqx{\end{displaymath}}
\def\beql{\arraycolsep .5mm \begin{eqnarray}}
\def\eeql{\end{eqnarray}}
\def\zeile{\nonumber \\[2mm] }   %  neue zeile
\def\rf#1{(\ref{#1})}
\def\labelsec#1{\label{SEC-#1}}
\def\rfsec#1{\ref{SEC-#1}}

\newcommand{\Section}[1]{\section[#1]{#1}
                         \setcounter{equation}{0}
   \renewcommand{\theequation}{\arabic{section}.\arabic{equation}}}

\newcommand{\Appendix}[2]{\section*{Appendix #1: #2}
                          \addcontentsline{toc}{section}{#1: #2}
                          \setcounter{equation}{0}
   \renewcommand{\theequation}{#1.\arabic{equation}}}

\def\d{{\rm d}}                    %Integrationsmass
\def\x{{\bf x}}                    %Ort x
\def\y{{\bf y}}                    %Ort y
\def\deltaxy{{\delta (\x , \y )}}  %
\def\Tr{{\rm Tr}}                  %

\def\mod{\ {\rm mod}\ }
\def\eins{{\sf \bf 1}}                 % Einheitsmatrix
\def\R{{\bf R}}                    % Reelle Zahlen
\def\rpar{\Lambda}                     % Regularisierungsparameter

\def\bareps{\bar \epsilon}

\def\ft#1#2{{\textstyle{{#1}\over{#2}}}}
\def\a{\alpha}
\def\b{\beta}
\def\g{\gamma}\def\G{\Gamma}
\def\D{\Delta}
\def\e{\epsilon}

\def\T{\Theta}
\def\l{\lambda}
\def\m{\mu}
\def\f{\phi}
\def\n{\nu}

\def\s{\sigma}\def\S{\Sigma}

\def\ve{\varphi}
\def\bx{\bar \xi}

\def\pois#1#2{ \{ #1 , #2  \}}
\def\komm#1#2{\big\lbrack #1 , #2 \big\rbrack }
\def\dirac#1#2{ \pois{#1}{#2}_* }

\def\euo#1#2{   e_{#1} ^{\;#2}  }
\def\eou#1#2{   e^{#1} _{\;#2}  }
\def\eoo#1#2{   e^{#1#2}  }
\def\euu#1#2{   e_{#1#2}  }
\def\Ao#1#2{   A_{#1} ^{\;#2}  }
\def\Au#1#2{   A_{#1#2}  }
\def\Euo#1#2{   E_{#1} ^{\;#2}  }

\def\neu{\prime}   % Kennzeichen fuer neues A,F, etc.
\def\AAo#1#2{  A_{#1}^{\neu   #2}  }

\def\Bo#1#2{  B_{#1}^{\; #2} }

\def\epuls#1#2{  {p_{#1}}^{#2}    }
\def\Apuls#1#2{  {\mit \Pi_{#1}}^{#2}    }

\def\V    {  {\cal V} }
\def\Vinv {  \V^{-1}  }
\def\VdV_#1{  \Vinv \partial_{#1}  \V }
\def\VDV_#1{  \Vinv D_{#1}  \V }
\def\ch#1{{\chi^{\dot #1}}}
\def\bch#1{{\bar \chi^{\dot #1}}}
\def\blam#1{{\bar \lambda_{\dot #1}}}
\def\JAB{{\cal J}_{\dot A \dot B}}

\def\eps{\varepsilon}      % epsilon-Tensor ( im Gegensatz zu
\def\Lag#1{{\cal L}^{(#1)}}
\def\Wirkung{S}
\def\Lagrange{{\c L}}
\def\Pexp{{\c P} \, \exp\, }

\def\Gam#1#2{ \Gamma^{ #1}_{ #2} } % Gamma-Matrizen fuer SUSY

\def\L{L}            % Lorentz
\def\T{T}            % Eichgruppe
\def\D{{\c D}}       % Diffoe             CONSTRAINTS
\def\H{{\c H}}       % Hamilton
\def\S{{\c S}}       % Susy
\def\K{{\c K}}       % H+\gamma H_k

\def\k{{ c}}     % Kurve
\def\J{{\c J}}       % komplexe Struktur fuer Fermionen

\def\c#1{{\cal #1}}
\def\suco#1{{\hat{#1}}}       % superkovariante Terme
\def\dens#1{{\widetilde #1}}     % gedichtet
\def\oper#1{{\widehat #1}}       % Quantenoperator
\def\dotbar#1{ \dot{\bar #1} }

\def\intd^#1#2{\int \! \d^#1 #2 \ } % \intd^2 \x erzeugt Integral

\def\deldel#1/#2/{\frac{\partial #1}{\partial #2}}
\def\deltadelta#1/#2/{\frac{\delta #1}{\delta #2}}

\def\grp#1({{\rm  #1}(}      % \grp SO(N)  fuer Gruppe
\def\alg#1({{\bf  #1}(}      % \alg so(N)  fuer Algebra

\def\f#1#2#3{ {f_{#1#2}}^#3 } %  Strukturkonstanten

\def\Euo#1#2{  {E_{#1}}^{#2}  }

\begin{document}
\setcounter{page}{0}
\thispagestyle{empty}
\begin{flushright} DESY 93-073 \\
                   gr-qc/9306018  \end{flushright}
\vspace*{1cm}
\begin{center}
{\LARGE \sc Canonical Quantum Supergravity in Three-Dimensions}\\
 \vspace*{1cm}
        {\sl H.-J. Matschull and H. Nicolai}\\
 \vspace*{6mm}
         II. Institut f\"ur theoretische Physik \\
          Universit\"at Hamburg \\
          Luruper Chaussee 149 \\
          22761 Hamburg \\
          Germany \\
 \vspace*{6mm}
June 15, 1993\\
\vspace*{1cm}
\begin{minipage}{11cm}\small
We discuss the canonical treatment and quantization of matter
coupled supergravity in three dimensions, with special emphasis
on $N=2$ supergravity. We then analyze the quantum constraint
algebra; certain operator ordering ambiguities are found
to be absent due to local supersymmetry. We show that the
supersymmetry constraints can be partially solved by a functional
analog of the method of characteristics. We also consider
extensions of Wilson loop integrals of the type previously found
in ordinary gravity, but now with connections involving the bosonic
and fermionic matter fields in addition to the gravitational
connection. In a separate section of this paper, the canonical
treatment and quantization of non-linear coset space sigma models
are discussed in a self contained way.
\end{minipage}
\end{center}
\newpage

\Section{Introduction}
\labelsec{intro}

The search for solutions of the Wheeler DeWitt equation \cite{WDW}
is one of the key issues of present day research in quantum gravity
(for a recent review and many further references, see e.g.
\cite{Isham}). Unfortunately, progress has been severely hampered by
technical problems, most notably the fact that the Wheeler DeWitt
equation is a non-polynomial functional differential equation
that is even difficult to {\it define} properly. The equation can
be substantially simplified by retaining only a finite number of
of degrees of freedom and thereby converting it into an ordinary
partial differential equation (which is still not
not easy to solve); this is the so-called
``mini-superspace approximation", see e.g. \cite{Isham} for
further explanations. Another, and perhaps more promising
attempt to come to grips with the Wheeler DeWitt
equation, which does
not involve any mutilation of the physical degrees of freedom, is
based on Ashtekar's new variables (see \cite{Ashtbook} for a recent
summary and many references). The main advantage of this approach,
which so far works only in three and four space time dimensions,
is that the canonical constraints become polynomial, which in turn
facilitates the search for solutions. Indeed, it is then possible
to construct formal solutions to all the constraints of pure quantum
gravity in four dimensions \cite{JacSmol, RovSmol}. At a kinematical
level, one can also incorporate matter in such a way that the
constraints remain polynomial; however, little progress has been
made so far in extending the results of \cite{RovSmol} to matter
coupled theories, but for special cases generalizations of
the Wilson loop variables may be constructed \cite{Matschull}.
                                        Furthermore,
it is not easy to see what has become of the singularities of
perturbative quantum gravity in this approach.
As a consequence, it is far from clear how the requirement of
quantum mechanical consistency could possibly affect matter
couplings in this approach, whereas experience with
string theory \cite{GSW} and $2d$ gravity \cite{GM}
would make us expect such consistency requirements to impose
stringent constraints on the allowed theories. In our opinion,
the inclusion of matter couplings and their proper treatment beyond
the purely kinematical aspects remains a major open problem.
The present work constitutes an attempt to address this problem in
the context of three dimensional supergravity.

This paper, then, deals with the canonical quantization
of matter coupled supergravities in three dimensions. It is based
on and considerably extends our previous results \cite{Nic91, MN},
where mostly classical aspects were studied. In section 2, we review
pure (topological) supergravities, which exist for any number
of local supersymmetries; this section will also serve to set up our
notations and conventions (see also \cite{MN}). The canonical treatment
of non-linear sigma models is discussed in section 3. Since the results
described there might also be of interest in other contexts,
and because the literature
on this topic seems to be scarce (see \cite{Nic91} for the canonical
formulation of $N=16$ supergravity and \cite{Forger} for a
discussion of flat space sigma models), we have aspired to make
this section self contained as far as possible. Section 4 is devoted
to a detailed study of the $N=2$ theory, which represents
the simplest non-trivial example of a locally supersymmetric
theory with matter couplings in three dimensions. Since the
generalization of these results to $N>2$ is to a large extent
straightforward, we have relegated the discussion of the higher $N$
theories to an appendix, where we explain the
redefinition of the gravitational connection required for the
decoupling of the phase space variables. A central part of this paper
is section 5, where we quantize the $N=2$ theory and analyze
its quantum constraint algebra. In particular, we will find that
at least some of the operator ordering ambiguities present in the
bosonic theories disappear due to local supersmmetry.
Unfortunately, apart from the trivial solution $\Psi =1$, we have so
far not been able to find solutions to all of the constraints.
Nonetheless, we can report some partial progress in this direction
by demonstrating that at least one half of the supersymmetry
constraints can be solved by a functional analog of the method
of characteristics; this requires the exponentiation of an
infinitesimal local supersymmetry transformation to a finite
transformation. Furthermore, we discuss a class of partial
solutions based on Wilson loop integrals over a connection
constructed out of the gravitational fields {\it and} the
matter fields,
which can be regarded as a ``supercovariant" extension
of the Wilson loop functionals considered in \cite{JacSmol}.

As is well known, supergravity theories are generally characterized
by rather complicated Lagrangians with non-polynomial scalar
self-interactions and quartic fermionic terms. Readers might
therefore wonder why one should choose to study them rather than
models with simpler matter couplings such as scalar or Dirac fields
without self-interactions. One of the main reasons why we prefer
these models over simpler ones is the {\it geometrical structure}
that is always present in the matter sectors of supergravity theories
and that is at the origin of their ``hidden symmetries" \cite{CJ}.
We believe that these symmetries may eventually play an important
role in improving our understanding of the matter coupled Wheeler
DeWitt equation for the following reason. Associated with the hidden
symmetries, there are non-trivial observables (or conserved charges)
in the sense of Dirac, which act on the space of solutions of the
quantum constraints. These symmetries may therefore be interpreted
as ``solution generating symmetries" for the Wheeler DeWitt
equation. An intriguing aspect is the emergence
of infinite dimensional symmetries acting on the space of classical
solutions of the gravitational field equations in the reduction to
{\it two} dimensions \cite{Geroch} (for more recent developments,
see \cite{Schlad}). If the theories could be quantized
in a way compatible with these symmetries, the Wheeler DeWitt
equation would become integrable in this reduction.

The fact that pure gravity in three dimensions is much easier to
quantize than theories of gravity in higher dimensions has been fully
appreciated only relatively recently, although classical aspects
(absence of gravitational excitations, i.e. gravitons,
in empty space, conical
singularities at the locations of matter point sources, etc.) have
been understood for a long time \cite{DesJacHoo}.
Since Einstein's action is superficially non-renormalizable in
three dimensions, the theory was for a long time thought to make no
more sense as a quantum theory than gravity in four dimensions. The
discovery that the quantum theory can be solved exactly came thus as
quite a surprise \cite{Witten88} (see also \cite{Martin} for further
studies of the quantum theory). An important ingredient in that
work was the reformulation of Einstein's theory as a Chern Simons
gauge theory. Here, we will, however, not make use of this
formulation, but rather adopt
an alternative and equivalent version based on
\cite{Asht et al}, which is a direct extension of Ashtekar's formalism
to three dimensions, and which provides an alternative route to
solving the quantum theory\footnote{We note that there is no reality
constraint on Ashtekar's variables in three dimensions unlike in four
dimensions \cite{Bengtsson}. However, this feature, which may be viewed
as another virtue of three dimensions, is lost when gravity is
coupled to fermionic matter, as we will explain in section 4
and the appendix.}. Both formulations in an essential
way exploit the fact that pure gravity in three dimensions is a
topological theory, whose physical phase space is related to the
moduli space of flat $\grp SL(2,\R)$ connections and hence
{\it finite}-dimensional for each genus. This result obviously
relies on the use of the gravitational (or spin) connection as the
primary canonical variable and would be much more difficult to obtain
in the usual metric formulation of quantum gravity.
Similar statements apply to pure supergravity in three dimensions,
which has been discussed extensively in \cite{MN}, where a solution
to the quantum constraints of pure $N=1$ supergravity has been
presented, and in \cite{oldWitten}, where a Chern Simons formulation
has been used. A common feature of the topological
theories is the existence of a complete set of observables in the
sense of Dirac, based on Wilson loops with or without dreibein and
gravitino insertions. By means of these observables, the solutions
to the quantum constraints can be obtained by applying the
observables to a suitable ``vacuum functional".

There are several reasons for studying locally supersymmetric theories
rather than non-supersymmetric ones. Local supersymmetry leads to a
constraint which can be thought of as the square root of the
Wheeler DeWitt
constraint, and is related to it in the same way as the Dirac equation
is related to the Klein Gordon equation (as was first observed in
\cite{Teitelboim}). However, due to the technical complexities,
the early papers on canonical supergravity \cite{DKS}
make no attempt at exploiting this idea, but content themselves with
setting up the canonical formalism and discussing the classical
constraint algebra in terms of Poisson (or Dirac) brackets. The
first investigation of the quantum theory appears to be \cite{D'Eath1},
where the metric formulation is utilized. More recently, there have
been several treatments of canonical quantum supergravity in the
mini-superspace approximation \cite{Graham}. A well
known feature of supersymmetric theories is the absence of certain
short distance singularities. From the analogy with the so-called
non-renormalization theorems of perturbative supersymmetric quantum
field theories \cite{Wess} and explicit calculations in
perturbative quantum supergravity \cite{PvN}
one would expect local supersymmetry to mitigate (if not
eliminate) the singularities occurring in the canonical constraint
operators as well, and thereby eliminate some of the
operator ordering ambiguities that afflict the canonical treatment
of non-supersymmetric theories \footnote{For instance, the solutions
of \cite{JacSmol, RovSmol} are based on the prescription that all
functional differential operators should be moved to the right.
If one chooses the opposite operator ordering prescription, one obtains
very different, and presumably inequivalent solutions \cite{Brueg}
(since the solutions of \cite{Brueg} require a non-vanishing
constant in contrast to \cite{RovSmol}).}.
In section \rfsec{quant}, we will show that this is indeed the case
for the constraint generators of supergravity. In particular, one
of the supersymmetry generators is a first order functional
differential operator and manifestly free of ordering ambiguities
and short distance singularities, while the other is a second
order operator, and the absence of ordering ambiguities is due to a
non-trivial cancellation. These results provide a first
glimpse as to what a non-perturbative non-renormalization
theorem in canonical quantum supergravity might look like.
However, it must be stressed that the question of non-perturbative
divergences and operator singularities cannot be resolved until
a scalar product in the space of physical states has been found.

While writing this paper, we received three preprints dealing with
canonical quantum $N=1$ supergravity in four dimensions.
\cite{Sano} discusses a supersymmetric extension of the
solution found in \cite{Brueg} (with non-vanishing cosmological
constant) and makes use of the Ashtekar formulation. \cite{D'Eath2}
is based on the metric formulation and proposes a solution very
similar to the Hartle-Hawking wave functional; see, however,
\cite{Page} for a criticism of this ansatz.

\Section{Pure Gravity and Supergravity in Three Dimensions}
\labelsec{pure}

The geometrical background for both pure and matter coupled
supergravity is characterized by a general three-dimensional space
time manifold, which is parametrized by local coordinates
$x^\m ,y^\m , \dots$. We use Greek letters $\m , \n , ...=0,1,2$ for
curved indices in three dimensions and Latin letters
$a,b,...$ for tangent space indices transforming under the local
Lorentz group $\grp SO(1,2)\cong \grp SL(2,\R)$. With $\euo \m a$ the
usual dreibein, the space time metric is given by $g_{\m \n} =
\euo \m a  \euo \n b \eta_{ab}$; it has has signature $(-++)$.
The Levi-Civita tensor with flat indices is defined by
$\e^{012} = -\e_{012} = +1$; it is related to the Levi-Civita
tensor density by
$\eps^{\mu\nu\rho} = e \euo \mu a \eou \nu b \eou \rho c \e_{abc}$.
Instead of the usual (first order) spin connection $\omega_{\m bc}$,
it is advantageous to use the dual connection
\beq
{A_\mu}^a = - \ft12 \eps^{abc} \omega_{\mu bc}
\eeq
in terms of which the Lorentz (i.e. $\grp SO(1,2)$) covariant
derivative acting on a three-component vector $V^a$ reads
\beq
D_\m V_a = \partial_\m V_a - \e_{abc} {A_\m}^b V^c
\eeq
The use of ${A_\m}^a$ rather than $\omega_{\m bc}$ simplifies the
canonical treatment considerably; in fact, as a canonical variable,
this field is the direct analog of Ashtekar's variable in three
dimensions. The field strength of the connection ${A_\m}^a$ is related
to the Riemann tensor by
\beq
F_{\mu \nu a} =
\partial_\mu A_{\nu a} - \partial_\nu  A_{\mu a}
 - \eps_{abc} {A_\mu}^b {A_\nu}^c
   = - \ft12 \eps_{abc} {R_{\mu \nu}}^{bc}
\label{3Feldstaerke}\eeq
so that Einstein's action becomes
\beq
S = \ft14 \int \d^3 x \, e \, R =
 \ft14 \int \d^3 x \eps^{\mu \nu \rho } {e_\mu}^a F_{\nu \rho a}
\label{3Einsteinwirkung}
\eeq

To introduce fermions, we make use of the real $\g$-matrices
$\g_0 = i\s_2$, $\g_1=\s_1$ and $\g_2=\s_3$, which satisfy
\beq
\gamma_a \gamma_b = \eta_{ab} {\bf 1} - \eps_{abc} \gamma^c
\eeq
The matrices $\g^a$ generate the group $\grp SL(2,\R)$ (the covering
group of the Lorentz group $\grp SO(1,2)$); because this group is
real, a reality constraint is not necessary unlike in four
dimensions \footnote{Our conventions regarding spinors are as
follows. The charge conjugation matrix is $C=\gamma^0$ and obeys the
usual properties $C^T=-C$ and $(C \g^a)^T =+ C \g^a$.
Majorana spinors satisfy $\bar \chi = \chi^T C$.
Later on, we will make use of the Fierz identity
\beqx
\bar \chi \ve \, \bar \l \psi  = - \ft12 \bar \l \ve \, \bar \chi \psi
 - \ft12 \bar \l \g^a \ve \, \bar \chi \g_a \psi
\eeqx
for anticommuting spinors $\chi , \ve , \l$ and $\psi$.
The underlying completeness
relation can also be expressed directly in terms of $\g$-matrices
\beqx
   \gamma^a_{\alpha\beta} \gamma_{a\gamma\delta}
     = - \delta_{\alpha\beta} \delta_{\gamma\delta}
      + 2 \delta_{\alpha\delta} \delta_{\gamma\beta}
\eeqx }.
The Lorentz covariant derivative on a spinor $\e$ reads
\beq
D_\mu  \epsilon = \left( \partial_\mu  + \ft12 \gamma_a {A_\m}^a
                       \right) \epsilon
\label{3DDefinition}
\eeq

In addition to the dreibein ${e_\m}^a$ and the connection field
${A_\m}^a$, we need $N$ gravitino fields $\psi_\m^I$, where
$I,J,...=1,...,N$. The Rarita Schwinger action in three
dimensions reads
\beq
    \Wirkung = \ft12 \intd^3 x
       \eps^{\mu\nu\rho} \bar \psi_\mu^I D_\nu \psi_\rho^I , \ \ \ \
       D_\nu \psi_\rho^I := \nabla\!_\nu \psi_\rho^I +
      \ft12 {A_\m}^a  \gamma_a \psi_\rho^I.
\label{SGR-Rarita-Schwinger}
\eeq
The sum of \rf{SGR-Rarita-Schwinger} and \rf{3Einsteinwirkung}
is invariant under the local supersymmetry transformations
\beq
    \delta_\epsilon \euo \mu a = \bar\epsilon^I \gamma_a \psi^I_\mu,
     \ \ \ \
    \delta_\epsilon \psi_\mu^I = D_\mu \epsilon^I,
\label{SGR-susy-trafo}
\eeq
It may seem curious that the combined action is supersymmetric for
arbitrary $N$, but this can be understood by noting that the
topological bosonic and fermionic degrees of freedom need {\it not}
balance in a supersymmetric theory unlike the propagating degrees
of freedom. The fact that the above theories are topological is
straightforward to verify. Namely, varying \rf{3Einsteinwirkung}
with respect to the dreibein we immediately deduce that the
field strength $F_{\mu \nu a}$ must vanish \footnote{The variation
with respect the connection
${A_\mu}^a$ tells us that the covariant derivative of the dreibein is
equal to a fermionic bilinear (torsion); this equation can be solved
for the connection in terms of the dreibein and the fermions
(``second order formalism" \cite{PvN}).} (note that, in three
dimensions, the Rarita Schwinger action is independent of the
dreibein and therefore does not contribute to this variation).
Hence, the connection ${A_\m}^a$ is pure gauge, at least
locally. However, ${A_\mu}^a$ may still be
non-trivial in that there may not exist a globally defined
function $g(x) \in \grp SL(2,\R)$ such that $A_\m \equiv
\ft12 {A_\mu}^a \g_a = g^{-1} \partial_\m g$.
Similar conclusions hold for the gravitino fields $\psi^I_\mu$.
The Rarita Schwinger equation $\epsilon^{\mu \nu \rho}
D_\nu \psi_\rho^I  = 0$ implies that $\psi_\mu^I$, too,
is locally pure gauge: we can always find a locally defined spinor
$\phi^I$ such that $\psi_\m^I = D_\m \phi^I $ (of course, this is only
true if $F_{\m \n} (A) =0$). Again, an obstruction only
arises if the spinor $\phi^I $ cannot be defined globally.

As is customary, for the canonical treatment\footnote{Standard
references on the canonical formulation of gravity are
\cite{Dirac, ADM, MTW}. For a general discussion of constrained
Hamiltonian systems, see \cite{Dirac, HRT}.} we will assume the space
time manifold to be the product of a spatial two-dimensional
hypersurface and the real line, parametrized by
the time coordinate $t$ \cite{Witten88}.
Derivatives with respect to $t$ are
denoted by a dot. Local coordinates on the spacelike
hypersurface will be denoted by $x^i, y^i,...$, or
simply just by bold face letters $\x ,\y ,...$, so that the
three-dimensional coordinates decompose as $x^\m = (t, x^i )$, etc.
There is a corresponding split of the three-dimensional
curved indices $\m , \n , \dots$ into a time index $t$ and spatial
indices $i,j, \dots$, so that $\m =(t,i)$, etc.
The Levi Civita tensor density splits as $\e^{tij} = \e^{ij}$, where
$\e^{ij} = \e_{ij}$ is the tensor density on the spacelike manifold.

Finally, we explain the canonical decomposition of the dreibein and
the metric. With $e={\rm det}\, {e_\m}^a$, we define the lapse and
shift variables by \cite{ADM}
\beq
    n := e g^{tt} , \ \ \ \  n^i := g^{ti}/g^{tt} .
\eeq
The dreibein is thus parametrized by the Lagrange multipliers $n$ and
$n^i$, and the remaining six components $\euo ia$, not all of which are
physical phase space degrees of freedom since three of them can be
eliminated in principle by local Lorentz rotations. The metric on the
spatial hypersurface, its inverse and determinant are then given by
\beq
   h_{ij} = g_{ij}  , \ \ \ \
   h^{ij} = g^{ij} - e^{-1} n n^i n^j , \ \ \ \
   h = - e n .
\eeq
The following polynomial functions of the dreibein components
$\euo ia$ will turn out to be useful
\beq
   \dens h^{ij} = h h^{ij} = \eps^{ik} \eps^{jl} g_{kl}, \ \ \ \
    e\eoo ta = - \ft12 \eps^{abc} \eps^{ij} \euu ib \euu jc ,
\eeq
Furthermore, it is convenient to employ a ``curved" basis for the
$\g$-matrices, which is given by $\g_i = \euo ia \g_a$
and $e\g^t = \ft12 \e^{ij} \g_i \g_j $; observe that
these, too, are {\it polynomial} functions of the dreibein
components. Given two three-vectors $X^\m$ and $Y^\m$, we have
\beql
e X^\mu Y_\mu & = & n X_n Y_n - n^{-1} \dens h^{ij} X_i Y_j,
    \zeile
e X_\mu \gamma^\nu \gamma^\mu Y_\nu & = &
     n X_n Y_n - n^{-1} \dens h^{ij} X_i Y_j \zeile & & {}
     + n^{-1}\,  \eps^{ij} X_i \,  e \gamma^t \, Y_j
     - \eps^{ij} ( X_n \gamma_i Y_j + X_i \gamma_j Y_n ),\zeile
  e \gamma^\mu X_\mu & = & e \gamma^t X_n
           + n^{-1} \eps^{ij} \, e \gamma^t\gamma_i X_j,
\eeql
The index $n$ here stands for the component normal to the spatial
hypersurface. This component is defined by $X_n = X_t + n^i X_i$ and
is related to $X^t = g^{t\m} X_\m$ by $n X_n = e X^t$.

\Section{Canonical Treatment of Non-Linear Sigma-Models}
\labelsec{sigma}

In the introduction, we have already mentioned the general result
that the bosonic sectors of (extended) supergravities are
governed by non-compact non-linear sigma-models. We will now
describe the canonical treatment of these models. Since the results
might also be useful in other contexts, we will temporarily ignore
all fields other than the scalars so as to make the discussion
self-contained.
Because our main interest is the application of the canonical
formalism to non-linear sigma models coupled to a non-trivial
gravitational background characterized by the metric $g_{\m \n}$,
we will, however, keep the dependence on the metric throughout;
the flat space models are then easily recovered by putting
$g_{\m \n}= \eta_{\m \n}$ everywhere.
Matter coupled supergravity theories in three dimensions have been
completely classified recently \cite{dWNT}. In contrast to
pure (topological) supergravity theories, which exist for any $N$,
the number of local supersymmetries is bounded by
$N\leq 16$ in the presence of matter couplings.
The matter sectors of these theories are described by non-linear
$\s$-models of the non-compact type \cite{CJ}, whose
target spaces become more and more restricted with
increasing $N$. More specifically, for three dimensional theories,
we have the following results \cite{dWNT}: for $N=1,2$ and 3,
the target manifolds $\c M$ are Riemannian, Kaehler and quaternionic,
respectively, whereas for $N=4$, the target space is locally a
product of two quaternionic manifolds associated with inequivalent
$N=4$ supermultiplets. Beyond $N=4$, only homogeneous (and, in fact,
symmetric) target spaces are allowed.

The standard sigma-model Lagrangian for an arbitrary Riemannian
target manifold $\c M$ is given by
\beq
\Lagrange = - \ft12
             e g^{\mu \nu} G_{mn} (\varphi ) \partial_\mu \varphi^m
        \partial_\nu \varphi^n
\label{SIGMALagrange1}
\eeq
where $\c M$ is parametrized by the coordinate fields $\ve^m (x)$ with
$m,n = 1,...,$ \\ ${\rm dim}\,  \c M$, and $G_{mn} (\ve )$ is a
Riemannian metric on $\c M$. Obviously, the main problem here
is posed by the non-linear interactions induced by the
geometrical form of this Lagrangian, and this problem also makes its
appearance in the canonical formalism. A first step in
resolving the difficulties is to select (canonical) quantities
that, despite their explicit dependence on the coordinate fields
$\ve^m$, transform as tensors under reparametrizations. Secondly,
we will see that a further simplification can be
achieved by utilizing tangent space tensors (tangent space, or just
``flat", target space indices will be designated by $A,B,\dots$).
Accordingly, we introduce a vielbein $\Euo mA (\ve )$ satisfying
\beq
G_{mn} (\ve )= \Euo mA (\ve ) \Euo nB (\ve ) \eta_{AB}
\label{SIGMAMetrik}
\eeq
where $\eta_{AB}$ is a flat metric in tangent space (which need not
be unique); in the following, we will freely use this metric to raise
and lower flat indices. We also define
\beq
P_\mu^A = \partial_\mu \varphi^m \Euo mA
\label{PmuDEF}
\eeq
where $\Euo Am$ is the inverse vielbein. The Lagrangian then
takes the simple form
\beq
\Lagrange = - \ft12 e g^{\mu \nu } P_\mu^A P_\nu^B \eta_{AB}
\label{SIGMALagrange2}
\eeq
This is also the form that appears in the supergravity
Lagrangians to be used later.

The canonical momenta, which are conjugate to the coordinate
fields $\ve^m$, are now easily calculated
\beq
p_m : ={ {\delta \Lagrange}\over {\delta \dot \ve^m}} =
   - e g^{t\mu }G_{mn}(\ve ) \partial_\mu \varphi^n
\label{SIGMAimpuls1}
\eeq
The basic Poisson brackets are given by
\beq
\pois  {p_m (\x )} {\varphi^n (\y)} = -\delta_m^n \deltaxy
\label{BasicPoisson}
\eeq
Although the momenta do transform properly under
reparametrizations (namely as vectors, i.e. elements of the tangent
space $T_\ve {\c M}$), the coordinate fields $\ve^m$ do not;
therefore, one must deal with non-covariant expressions at the
intermediate stages of every calculation if one uses these variables.
The Hamiltonian is given by
\beq
   H := \intd^2\x \big( p_m \dot \ve^m - \Lagrange \big)
\label{SIGMA-Hamilton}
\eeq
Canonical quantization will be awkward to carry out in terms of the
variables $\ve^m$ and $p_m$ due to operator ordering problems and the
concomitant short distance singularities (which may also spoil
general covariance in target space by ``anomalies"). In any case,
quantization will require a definite ordering prescription
for the operators involving the momenta $p_m$. Here, we find it
convenient to employ another set of canonical
variables and to perform the quantization
directly in terms of them rather than in terms of
of the original variables $\ve^m$ and $p_m$. This procedure defines
the quantum theory in an unambiguous way, as it corresponds to
a definite choice of operator ordering.

As our basic canonical variables, we choose the
``composite" quantities
\beq
P_A : = {{\delta \Lagrange} \over {\delta P_t^A}}  =
        \Euo Am (\ve ) p_m  \;\;\; , \;\;\;
P_i^A : =   \partial_i \varphi^m \Euo mA (\ve )
\label{SIGMAImpuls2}
\eeq
The Hamiltonian \rf{SIGMA-Hamilton} can also be obtained from
\beq
     H(P_A, \ve ) =
  \intd^2\x \big( P_A P_t^A  - \Lagrange \big)
\label{SIGMAHamilton2}
\eeq
The variables $P_A$, which we now regard as the momenta,
evidently correspond to an anholonomic basis in tangent space
(whereas the $p_m$ are like a coordinate basis). Our
choice is also motivated by the fact that the variables
\rf{SIGMAImpuls2} are precisely the ones which will appear in the
the supergravity constraints to be derived in later sections.

To compute the canonical brackets of $P_A$ and $P_i^A$, we employ the
basic Poisson brackets \rf{BasicPoisson}. Of course, it does not matter
at this point whether or not we use the original fields $\ve^m$
and $p_m$ for this purpose; afterwards, we can simply ``forget" how
the results were derived. A straightforward calculation yields
\beql
\pois {P_A(\x )} {P_B(\y )} &=&  {\Omega_{AB}}^C (\x ) P_C (\x )
  \, \deltaxy       \zeile
\pois {P_A(\x )} {P_i^B (\y )} &=&  \Big( \delta_A^B \partial_i -
   {\Omega_{AC}}^B (\x ) P_i^C (\x )\Big) \, \deltaxy   \zeile
\pois {P_i^A (\x )} {P_j^B (\y )} &=&  0
\label{Algebra}
\eeql
where ${\Omega_{AB}}^C:= 2{E_{[A}}^m {E_{B]}}^n  \partial_m {E_n}^C$
are the coefficients of anholonomy (by $\partial_m$, we denote
the derivative with respect to $\ve^m$). Here and in the remainder,
spatial derivatives $\partial_i$ will always be understood to act on
the {\it first} argument in the $\delta$-function (i.e. $\x$ in
\rf{Algebra}). The Poisson brackets
\rf{Algebra} will be regarded as the basic relations from now on.
If we parametrize phase space in terms of the variables $\ve^m$
and the momenta $P_A$, these brackets are reproduced by the
general formula
\beql
\pois fg & = & \intd^2\x   \Big(
              \Euo Am  \deltadelta f /  \ve^m(\x) /
                       \deltadelta g /  P_A(\x) /
         -    \Euo Am  \deltadelta g /  \ve^m(\x) /
                       \deltadelta f /  P_A(\x) / \zeile && \ \ \ \ \
  \ \ \ \ + {\Omega_{AB}}^C (\x ) P_C(\x) \deltadelta f /  P_A(\x) /
                               \deltadelta g /  P_B(\x) / \Big)
\label{LIE-Poisson-coordinates}
\eeql
where $f$ and $g$ are arbitrary functionals of $\ve^m$ and $P_A$.

The transition to the quantized theory is implemented by the
replacement
\beql
P_A (\x ) & \longrightarrow   &
\oper {P_A} (\x )  = - i \Euo A m \big( \ve (\x )\big) {\delta \over
    {\delta \ve^m (\x )}}   \zeile
P_i^A (\x ) &\longrightarrow  &
\oper {P_i^A} (\x ) =\partial_i \ve^m (\x )\Euo mA \big( \ve (\x )\big)
\label{OPDarstellung1}
\eeql
The ordering prescription implicit in this replacement ensures that
the relations \rf{Algebra} can be directly replaced by quantum
mechanical commutators (modulo factors of $i$), and the
geometrical structure of \rf{Algebra} is thus preserved.

At this point, not much more can be said if the target space $\c M$
is an arbitrary Riemannian manifold. For this reason and also
in view of the fact that the target manifolds of supergravity
are usually constrained by local supersymmetry to be of a very
special type, we will now make further assumptions on the
structure of $\c M$. The simplest possibility is to assume that the
target space is a group manifold, i.e. $\c M = G$ for some Lie
group $G$. Although the target spaces relevant to our investigation
are $not$ group manifolds in general, we discuss this case first
since all relevant formulas can be derived from it. This is because,
as we will explain below, we can formally treat the coset manifolds
occurring in supergravity on the same footing as group manifolds
if we add suitable gauge degrees of freedom.

For group manifolds, we assume the vielbein
\rf{SIGMAMetrik} to be a left invariant vector field; this means that
\beq
 \Euo Am (\tilde \ve (\ve )) = \Euo An (\ve )
   \frac{\partial \tilde \ve^m(\ve)}{\partial \ve^n}.
\label{LIE-left-invariance}
\eeq
where $\ve \rightarrow
\tilde \ve (\ve )$ is the diffeomorphism induced by left
multiplication. Then, the coefficients of anholonomy and the
flat metric are
given by the structure constants of $G$, viz.
\beq
{\Omega_{AB}}^C = - {f_{AB}}^C, \ \ \ \ \
\eta_{AB} = {f_{AC}}^D {f_{BD}}^C,
\label{OgleichF}
\eeq
where the structure constants $f_{ABC}$ are defined through the
commutation relations
\beq
\big[ Z_A , Z_B \big] = {f_{AB}}^C Z_C
\label{DefFABC}
\eeq
for the generators $Z_A$ of $G$. The vielbein $\Euo mA$ can be
explicitly computed by introducing a matrix representation
$\V = \V(\ve^m (x)) \in G $
\beq
\VdV_m = \Euo m A   Z_A
\eeq
where as before $\partial_m$ denotes the derivative with respect
to the coordinate field $\ve^m$.
{}From \rf{PmuDEF}, we get the identification
\beq
P_\m^A Z_A = \partial_\m \ve^m \, \VdV_m =  \VdV_\m
\label{PmuGruppe}
\eeq
The field theoretic model obtained in this way goes by the
the name of ``principal chiral model"; its Lagrangian is simply
obtained by substituting \rf{PmuGruppe} into
\rf{SIGMALagrange2}. From the results listed above,
we can immediately derive the relevant brackets by substituting
the structure constants for the coefficients of anholonomy; in
addition, we can determine the brackets between the momenta $P_A$
and the matrices $\V (\ve)$. The result is
\beql
  \pois{ P_A(\x) }{ P_B(\y) } &=& -\f ABC P_C \, \deltaxy, \zeile
  \pois{ \V(\x)  }{ P_A(\y) } &=& \V Z_A \, \deltaxy ,
\label{LIE-Poisson-matrix}
\eeql
As a check, we can recalculate the brackets between $P_A$ and $P_i^B$
(cf. \rf{Algebra}) from these formulas and \rf{PmuGruppe}.

The above formulas are not yet quite what we want, since the
relevant target spaces to be considered in the remainder are
coset spaces rather than group manifolds; however, the above
brackets will nonetheless prove useful in that they will enable
us to compute the relevant Poisson brackets for coset space sigma
models as well. As is well known, any symmetric space can be
represented as a coset $G/H$; in the case at hand, the group $G$ is
non-compact, and $H$ its maximally compact subgroup \cite{CJ}.
There are now two equivalent formulations. One either
parametrizes the manifold $\c M = G/H$ in terms of coordinates
$\ve^m$ with $m=1,..., {\rm dim} \, G/H$ as described above;
or one introduces extra coordinate fields $u^r (x)$ associated with
the subgroup $H$ (so that $r=1,..., {\rm dim} \, H$), in which
case the coordinates $(\ve^m , u^r )$ parametrize the whole
group $G$. If one uses only the physical fields $\ve^m (x)$,
part of the invariance under the isometry group $G$ is realized
non-linearly. In the second case, the invariance transformations
under the isometry group can be realized linearly, at the expense of
introducing an artificial gauge invariance necessary to remove
the unphysical degrees of freedom corresponding to the fields
$u^r (x)$. In the canonical formalism, this gauge invariance
will lead to constraints.

Since we prefer to make use of the second formulation, let us
introduce a matrix representation $\V = \V(\ve^m (x) , u^r (x) )\in G$.
To get rid of the unwanted degrees of freedom which are represented
by the fields $u^r (x)$, we postulate in addition that the
Lagrangian should be invariant under the transformations
\beq
\V (x) \longrightarrow g \V (x) h(x)
\label{VTrafo}
\eeq
with $g \in G$ and $h(x) \in H$, so that putting $u^r =0$,
we recover the description in terms of the physical fields $\ve^m$
(this gauge choice is sometimes referred to as the ``unitary gauge").
We split the generators $Z$ of $G$ into the generators
$X^\a$ of $H$ ($\a ,\b ,...= 1,...,r$) and the remaining coset
generators $Y^A$ \footnote{We hope that the dual use of the indices
$A,B,...$ will not cause confusion; they label either all group
generators as in \rf{DefFABC} or just the coset generators as here.};
the structure constants are decomposed accordingly.
For a {\it symmetric} space, we have $f_{ABC}= f_{\a \b C} = 0$,
and \rf{DefFABC} reads
\beq
\big[ Y^A , Y^B \big] = {f^{AB}}_\g X^\g  \;\; , \;\;
\big[ X^\a , Y^B \big] = {f^{\a B}}_C Y^C  \;\; , \;\;
\big[ X^\a , X^\b \big] = {f^{\a \b}}_\g X^\g
\eeq
To write down the Lagrangian, we decompose the Lie algebra valued
expression $\VdV_\m$ according to
\beq
\VdV_\m = P_\m^A Y_A + Q_\m^\a X_\a
\label{VDV}
\eeq
The Lagrangian is then again given by \rf{SIGMALagrange2}. Note,
however, that the sum over $A$ now runs only over the coset
generators. As a consequence, the fields $Q_\m^\a$ do not appear
the Lagrangian (however, they do couple to the fermionic fields
in its supersymmetric extension); they are just the gauge fields
required by local $H$ symmetry. So we see that it is the Lagrangian
that determines which degrees of freedom are physical and which are
not; we can convert the principal chiral model
into a coset space sigma model simply by omitting
those $P_\m^A$ corresponding to a subgroup of $H$ from the sum
\rf{SIGMALagrange2}. Of course, for a non-compact group $G$, there
is only one choice of the subgroup $H$ for which the Hamiltonian
is positive definite. If we define the canonical momenta by
\beq
P_A : = {{\delta \Lagrange} \over {\delta P_t^A}}  \;\;\; , \;\;\;
Q_\a := {{\delta \Lagrange} \over {\delta Q_t^\a}}
\eeq
the absence of $Q_t^\a$ from the Lagrangian immediately implies
the constraint $Q_\a = 0$; this must be interpreted as a weak
equality in accordance with the general theory of constraints
\cite{Dirac, HRT}. The Hamiltonian is now given by
\beq
     H(P_A, Q_\a , \ve , u) =
  \intd^2\x \Big( P_A P_t^A +  Q_\a Q_t^\a - \Lagrange \Big)
\eeq
We repeat that the main difference from the canonical point of view
between the principal chiral model and the coset space sigma model
characterized by this Hamiltonian is that the momenta $Q_\a$
corresponding to the subgroup $H$ have become constraints.
Nonetheless, the combined set of momenta $P_A$ and $Q_\a$ still obeys
the same Poisson brackets as before; consequently, we can read
off the result directly from \rf{LIE-Poisson-matrix}. So, we get
\beql
\pois{ P_A(\x) }{ P_B(\y) } &=& -\f AB{\g} Q_\g (\x )\, \deltaxy, \zeile
\pois{ Q_\a (\x )}{P_B(\y) } &=& -\f {\a}BC P_C (\x )\, \deltaxy ,\zeile
\pois{ Q_\a (\x )}{Q_\b (\y )}&=& -\f {\a}{\b}{\g} Q_\g (\x )\, \deltaxy
\label{ConstraintAlgebra}
\eeql
Furthermore,
\beql
  \pois{ \V(\x)  }{ P_A(\y) } &=& \V Y_A \, \deltaxy , \zeile
  \pois{ \V(\x)  }{ Q_\a (\y) } &=& \V X_\a \, \deltaxy
\label{KommutatorVPQ}
\eeql
which shows that the $Q_\a$ generate local $H$ transformations
on $\V$.

To construct an operator representation for the $P_A$ and $Q_\a$,
we could simply take over formula \rf{OPDarstellung1} with an
appropriate split of indices. The matrix $\Euo mA$ would then
have to be decomposed accordingly. The general parametrization
of $\V$ in terms of $\ve^m$ and $u^r$ adopted so far would,
however, lead to formulas which are somewhat unwieldy for practical
calculations, as the physical and unphysical degrees are difficult
to disentangle. For this reason, we choose a slightly different
parametrization in terms of which the constraints are easier to solve.
Locally, we can always assume that the matrix
$\V (\ve , u)$ can be written in the form
\beq
\V (\ve ,u) = \V_0 (\ve ) h(u)
\eeq
Then a straighforward calculation shows that
\beql
\VdV_m &=& Q_m^\a (\ve ,u) X_\a + \Euo mA (\ve ,u) Y_A  \zeile
\VdV_r &=& \Euo r{\a} (u) X_\a
\eeql
where we have expanded the right hand side in terms of the subgroup and
the coset generators, thereby defining the various submatrices.
Consequently, the vielbein on $G$ (which is a ${\rm dim} \, G \times
{\rm dim} \, G$ matrix) is triangular
\beq
 {\rm Vielbein}   =
  \pmatrix{ \Euo mA (\ve ,u) &  Q_m^\a (\ve ,u)  \cr
            0  & \Euo r{\a}(u)  \cr}
\eeq
The advantage of this parametrization is that the
${\rm dim}\, G/H \times {\rm dim} \, G/H$ matrix $\Euo mA (\ve ,u)$
can be identified with the vielbein on $\c M = G/H$ after a
$u$-dependent tangent space rotation. The inverse vielbein on $G$
is given by
\beq
 {\rm Inverse}\, {\rm Vielbein}   =
  \pmatrix{ \Euo Am (\ve ,u) &  -\Euo Am Q_m^\b \Euo {\b}r (\ve ,u) \cr
            0  & \Euo {\a}r (u)  \cr}
\eeq
and $\Euo Am$ can be identified with the inverse vielbein on $G/H$ up
to a $u$-dependent $H$ rotation. Inserting these expressions into
\rf{OPDarstellung1} and relabeling indices, one arrives at
the following operator representations
\beql
\oper {P_A} (\x )&=&
       i\Euo Am \left( {\delta \over {\delta \ve^m (\x )}}-
   Q_m^\b \Euo {\b}r {\delta \over {\delta u^r (\x ) }}  \right)\zeile
\oper {P_i^A} (\x ) &=& \partial_i \ve^m \Euo mA    \zeile
\oper {Q_i^\a} (\x ) &=&
      \partial_i \ve^m Q_m^\a + \partial_i u^r \Euo r{\a}
\label{OPDarstellung2}
\eeql
The constraint $Q_\a$ is realized by the operator
\beq
\oper {Q_\a} (\x ) =
              i\Euo {\a}r {\delta \over {\delta u^r (\x ) }}
\eeq
and depends only on the gauge degrees of freedom. Observe that
the momentum operator $\oper {P_A}$ can be viewed as a
connection on the principal fiber bundle $G \rightarrow G/H$
with base space $G/H$ and fiber $H$ (it defines a ``horizontal
subspace" of $T_{\{\ve,u\} } G$ at each point); note, however,
that we are dealing with {\it functional}, not ordinary
derivatives here.  We recall that in the quantized theory, any
physical wave functional $\Psi [\ve ,u ]$ must satisfy
$\oper Q_\a \Psi =0$; with the above parametrization, this is simply
solved by $\Psi = \Psi [\ve ]$. We emphasize however, that the
$u$-dependence of $\oper P_A$ cannot be dropped since otherwise the
constraint algebra \rf{ConstraintAlgebra} would not be obeyed.

{}From \rf{VTrafo} it follows that $G$ acts as a group of isometry
transformations on the target space $\c M = G/H$. The associated
charge density $\c J (\x )$ (which is a matrix with values in
the Lie algebra of $G$) is obtained by sandwiching the momenta and
constraints between the matrix $\V$ and its inverse. Thus
\beq
{\cal J}(\x ) = \V \left( P_A Y^A + Q_\a X^\a  \right) \Vinv
\eeq
The charges
\beq
{\cal Q} = \int \d^2 {\x} \, {\cal J} (\x )
\label{Charge}
\eeq
constitute the canonical generators of the isometry group,
and generate the isometry transformations on the fields,
as can be verified from the relations \rf{KommutatorVPQ}.
Incidentally, this formula also remains valid in supergravity
since the rigid group $G$ does not act on the fermions, or
only via induced $H$-rotations. The above expressions for the charge
differs from the one given in \cite{Nic91} by the constraint
generator of $H$ gauge transformations, which precisely removes
the fermionic bilinears in the formulas given there.

To conclude this section, we give another and equivalent
representation of the operators \rf{OPDarstellung1} and
\rf{OPDarstellung2}, to be used in section 5.
When working with a concrete matrix
representation $\V$, we can regard the elements of $\V$
as independent fields. To be sure, we would then have to introduce
second class constraints to ensure that $\V$ remains an element of
the group $G$. However, there is no need to enter into the details of
this construction here, as long as $\V$ is always understood to be
an element of the group $G$ in all formulas below.
Let us simply assume that the Lagrangian is given as a
function of the matrix field $\V (x)$ and its inverse $\Vinv (x)$ as
well as its ``derivatives" $P_\m^A (x)$. Similarly, the physical states
are assumed to be represented by wave functionals $\Psi$ which
depend on $\V$, $\Vinv$ and their spatial derivatives. On any
such wave functional, we define the momentum operators
$\oper {P_A}$ through their action on $\V$ and $\Vinv$, which is
given by $\oper P_A \V := i \V Z_A$ and $\oper {P_A} \Vinv : =
-i Z_A \Vinv $. It follows immediately that
$\komm {\oper P_A}{\oper P_B} = i\f ABC \oper P_C $.
Defining the matrix valued derivative operator $\delta/\delta \V$
by $(\delta/\delta \V)_{pq} := \delta/\delta \V_{qp}$, the
operator representation for the momenta becomes
\beq
   \oper P_A = i\, \Tr \Big( \V Z_A \deltadelta/ \V / \Big),
\label{LIE-quantum-operator}
\eeq
Since the matrices $Z_A$ generate the Lie algebra of $G$,
the action of this operator is tangent to the submanifold defined
by the group $G$ in the space of all matrices $\V$, and hence does
not depend on how we define the functional $\Psi [\V ]$ away from it.
If we are dealing with a coset space sigma model,
the same remarks as before apply; we simply have to split the group
indices into subgroup and coset indices, and the momenta corresponding
to the subgroup become constraints. However, the solution
to the constraint $\oper Q_\a \Psi = 0$ apparently cannot be cast
into a simple form in this representation. Finally, we note
the simple expression for the quantum mechanical charge operator
\rf{Charge} in this representation; it is
\beq
{\c Q} = {\c Q}_A Z^A , \ \ \ \ \
{\c Q}_A =  \int \d^2 \x \, \Tr \Big( Z_A \V \deltadelta / \V / \Big).
\label{chargerep}
\eeq
(If we are dealing with a coset space sigma model, the generators
are ${\c Q}_A$ and ${\c Q}_\a$). It is easy to check that $\c Q$
generates the global $G$-transformations acting from the left on
$\V$ according to \rf{VTrafo}. Furthermore,
\beq
\big[  \oper P_A \, , \, {\c Q} \big] = 0
\label{chargeconstraint}
\eeq
In particular, the global charges commute with the constraints
$\oper Q_\a$ for a coset space sigma model and are thus observables
in the sense of Dirac, or ``conserved charges".

\Section{The $N=2$ Theory}
\labelsec{N=2}

The simplest nontrivial example of matter coupled supergravity in
three dimensions is provided by the $N=2$ theory \cite{MSS, MN}.
We will here follow the presentation in \cite{MN}, where this model
has been described in great detail, and only summarize
its main features before moving on to the canonical formulation.
In addition to the gravitational degrees of freedom, the theory
contains two gravitinos $\psi^I_\m$ ($I=1,2$), two matter fermions
$\chi^1$ and $\chi^2$ and two (real) bosons that live on the
coset space $\grp SL(2,\R)/\grp SO(2)$. The fermions are real (Majorana)
spinors, which we will combine into complex (Dirac) spinors by
defining $\psi_\m = \ft1{\sqrt2} (\psi^1_\m  + i\psi^2_\m )$ and
$\chi = \ft1{\sqrt2} (\chi^1 + i\chi^2 )$. If the $\grp SL(2,\R)$
symmetry is linearly realized, the fermions transform only
under the gauge group $H$. The case of abelian $H$ is a little
peculiar as the relative normalization between the $H$ generators
and the coset (i.e. $G/H$) generators is not fixed, unlike
for non-abelian $H$. Consequently, the requirement of local
supersymmetry and $H$ invariance does not uniquely determine the
fermionic $\grp SO(2)$ charges in contrast to the theories with
$N\geq 3$  \footnote{This distinction between abelian and non-abelian
subgroup $H$ was first emphasized by B. de Wit (private
communication).}. Our charge assignments agree with those used
in our previous work \cite{MN} and coincide with the ones obtained
by dimensional reduction of $N=1$ supergravity in four
dimensions \cite{DZ, PvN} to three dimensions (but differ from the
ones that one would obtain from the Lagrangian given in \cite{dWNT}).
Thus the matter fermion $\chi$ has charge $+ \ft32$ and the
gravitino field $\psi_\mu$ has charge $-\ft12$; the $\grp SO(2)$
group can then be interpreted as the helicity group of the
four-dimensional ancestor theory.

As explained in the foregoing section, we will parametrize the
bosonic fields by a matrix $\V$ which takes values in the group
$\grp SL(2,\R)$. The unphysical degree of freedom
corresponding to the subgroup $H$ is removed by postulating
invariance under local $H$ transformations as in \rf{VTrafo}.
In accordance with the notation used in section \rfsec{sigma}, we denote
the generator of the $\grp SO(2)$ subgroup by $X$,
and the remaining generators by $Y^1$ and $Y^2$. Again, we find it
convenient to switch to a complex basis
$Y=\ft1{\sqrt2} (Y_1+iY_2) , Y^* = \ft1{\sqrt2} (Y^1-iY^2)$.
The $\grp SL(2,\R)$ commutation relations then read
\beq
   \komm{ X } { Y } =   2iY  , \ \ \ \
   \komm{ X } {Y^*} =  -2iY^* , \ \ \ \
   \komm{ Y   } { Y^* } =-2iX ,
\label{2-commutators}
\eeq
and formula \rf{VDV} becomes
\beq
   \VdV_\mu = P_\mu^* Y + P_\mu Y^* +  Q_\mu X,
\label{2-PQ-definition}
\eeq
Remembering our $\grp SO(2)$ charge assignments and \rf{3DDefinition},
we can immediately write down the fully covariant derivatives
on the spinors
\beql
   D_\mu \chi & := & \partial_\mu \chi
               + \ft12 \Ao \mu a \gamma_a \chi
               - \ft32 i Q_\mu  \chi   \zeile
   D_\mu \psi_\nu & := & \nabla\!_\mu \psi_\nu
               + \ft12 \Ao \mu a \gamma_a \psi_\mu
               + \ft12 i Q_\mu \psi_\nu, \zeile
   D_\mu \euo \nu a & := &  \nabla\!_\mu \euo \nu a
               - \eps^{abc} \Au \mu b \euu \nu c ,
\label{2-covariant derivatives}
\eeql
The Lagrangian of $N=2$ supergravity is then given by
\beq
\Lagrange = \Lag0 + \Lag1 + \Lag2
\label{N2Lagrange}
\eeq
where
\beql
   \Lag0 &=& \ft14 \eps^{\mu\nu\rho} \euo \mu a F_{\nu\rho a}
           + \eps^{\mu\nu\rho} \, \bar\psi_\mu D_\nu \psi_\rho,\zeile
   \Lag1 &=&{} - e g^{\mu\nu} P_\mu P_\nu^*
         + e \, \bar \psi_\mu \gamma^\nu \gamma^\mu \chi \, P_\nu^*
         + e \, \bar \chi \gamma^\mu \gamma^\nu \psi_\mu \, P_\nu
           \zeile &&{}
         - \ft12 e \, \bar \psi_\mu \gamma^\nu \gamma^\mu \chi \,
                 \bar \chi \psi_\nu
         - \ft12 e \, \bar \chi \gamma^\mu \gamma^\nu \psi_\mu \,
                 \bar \psi_\nu \chi,\zeile
   \Lag2 &=&{} -\ft12 e \, \bar \chi \gamma^\mu D_\mu \chi\,
               + \ft12 e  \, D_\mu \bar \chi \gamma^\mu \chi\,
               - \ft32 e \, \bar \chi \chi \, \bar \chi \chi.
\label{2-Lagrange}
\eeql
Apart from the contribution involving the gauge field $Q_\mu$,
$\Lag0$ is identical with the topological Lagrangian introduced
in section \rfsec{pure} (see \rf{3Einsteinwirkung} and
\rf{SGR-Rarita-Schwinger}) for $N=2$.
The full Lagrangian \rf{N2Lagrange} is invariant under the
local supersymmetry transformations
\beql
  \delta_\epsilon \psi_\mu  & = & D_\mu \epsilon
      + \ft12 \eps_{\mu\nu\rho} \gamma^\nu \epsilon \,
               \bar \chi \gamma^\rho \chi , \zeile
  \delta_\epsilon \euo \mu a & = &
           \bar \epsilon \gamma^a \psi_\mu
         - \bar \psi_\mu \gamma^a \epsilon , \zeile
  \delta_\epsilon \V & = &
          \bar \chi \epsilon \, Y + \bar \epsilon \chi Y^*,\zeile
  \delta_\epsilon \chi & = & \gamma^\mu \epsilon \, \suco P_\mu ,
\label{2-susy-trafo}
\eeql
where $\suco P_\mu := P_\mu - \bar \psi_\mu \chi$ is the
supercovariantization of $P_\mu$. We refrain from giving the
variation of the gravitational connection $\Ao\mu a$, as the
variation of the action under \rf{2-susy-trafo} is proportional
to the torsion equation, and therefore $\delta {A_\m}^a$ can be
chosen so as to cancel this contribution (see e.g. \cite{PvN}).
The torsion equation reads
\beq
    D_{[\mu} \euo {\nu]} a =
           \bar \psi_{[\mu} \gamma^a \psi_{\nu]}
           - \ft12 \eps^{abc} \euu \mu b \euu \nu c \, \bar\chi\chi
\label{2-torsion-equation}
\eeq

In the canonical treament of this model, one encounters the
following technical difficulty \cite{MN}. As it turns out,
the Dirac brackets between the components of
${A_i}^a$ do {\it not} vanish, but rather commute to
give a bilinear expression in the matter fermions $\chi$;
furthermore, the Dirac brackets between ${A_i}^a$ and $\chi$
do not vanish, either. This feature prevents straightforward
quantization via the replacement of phase space variables by
functional differential operators. It has already been pointed out
in \cite{MN} that, for the $N=2$ theory, the phase space variables
can be decoupled through a redefinition of the gravitational
connection by a fermionic bilinear\footnote{A very similar
redefinition is necessary in the metric formulation \cite{D'Eath1}.}.
In \cite{MN}, this redefinition
was performed {\it after} setting up the Hamiltonian formulation,
but, as we shall now demonstrate, it is much more convenient to do
so already at the level of the Lagrangian, since this will entail
substantial simplifications. For this purpose, we define a new
connection field ${A'_\m}^a$ by
\beq
   {A'_\m}^a :=  {A_\m}^a + \eps^{abc} \euu \mu b \,
           \bar \chi \gamma_c \chi
\label{2-new-connection}
\eeq
Observe that the fermionic bilinear is purely imaginary, and
hence the connection becomes {\it complex}.
The redefinition and the decoupling also work for the higher
$N$ theories, although the details are more involved and are
therefore explained in appendix A.

To see how this redefinition affects the Lagrangian, we now
substitute the new connection \rf{2-new-connection} into the
the above Lagrangian. For the gravitational curvature, one finds
\beq
     F_{\mu\nu a}  =
     F_{\mu\nu a}^\neu  - 2 \eps_{abc} D^\neu_{[\mu} \big(
         \euo {\nu]} b \, \bar \chi \gamma^c \chi \, \big)
        - 3 \eps_{abc} \euo \mu b \euo \nu c \, \bar\chi \chi
                                             \, \bar \chi \chi,
\eeq
Here, the prime on the covariant derivative indicates the replacement
of ${A_\mu}^a$ by ${A'_\mu}^a$. Because ${A'_\m}^a$ is complex,
the Dirac conjugate of $D_\mu^\neu \chi$ would involve
the complex conjugate connection; it is thus different from the
the derivative $D_\mu^\neu \bar\chi$ used in the above
equation, which is
defined by $D_\mu^\neu \bar\chi = \partial_\mu \bar\chi - \ft12
{A'_\m}^a \bar \chi \gamma_a + \ft32 i Q_\mu \bar \chi $. On the
other hand, $\Lag1$ does not contain ${A_\m}^a$ and is therefore
unchanged. In $\Lag2$ we may replace ${A_\m}^a$ by ${A'_\m}^a$, because
the difference vanishes by simple symmetry arguments.

Insertion of the new connection into $\Lag0$ yields
\beql
  \ft14 \eps^{\mu\nu\rho} \euo \mu a F_{\nu\rho a} & = &
  \ft14 \eps^{\mu\nu\rho} \euo \mu a F_{\nu\rho a}^\neu
  - \ft12 e \eou \mu a \, D_\mu^\neu (\, \bar \chi \gamma^c \chi \, )
  + \ft32 e \, \bar \chi \chi \, \bar \chi \chi,\zeile
  \eps^{\mu\nu\rho} \, \bar \psi_\mu D_\nu \psi_\rho & = &
  \eps^{\mu\nu\rho} \, \bar \psi_\mu D_\nu^\neu \psi_\rho
    - \ft12 \eps^{\mu\nu\rho}
     ( \, \bar \psi_\mu \gamma_\nu \chi \, \bar \chi \psi_\rho
     - \bar \psi_\mu \chi \, \bar \chi \gamma_\nu \psi_\rho \, ) ,
\label{2-Lgrav-inserted}
\eeql
where the first equation holds up to a total derivative only.
We observe two crucial results. First, most of the higher order
fermionic terms in the action are canceled by the redefinition
of the spin connection, which greatly simplifies the Lagrangian.
Secondly, the Einstein term now contributes to the Dirac term
in such a way that all derivatives on $\bar \chi$ disappear
from the action. This is rather fortunate, because otherwise, either
hermiticity would be lost (if all derivatives were defined with the
complex connection ${A'_\m}^a$, as we did above),
or we would have to introduce the complex conjugate connection
$({A_\m}^a)^*$, which would not be subject to
simple commutation relations (cf. \rf{2-Dirac-brackets} below)
and spoil the decoupling of bosonic and fermionic fields in the
canonical brackets.

The total Lagrangian in terms of the new connection is now the sum of
\beql
   \Lag0 &=& \ft14 \eps^{\mu\nu\rho} \euo \mu a F_{\nu\rho a}
           + \eps^{\mu\nu\rho} \, \bar\psi_\mu D_\nu \psi_\rho,\zeile
   \Lag1 &=& - e g^{\mu\nu} P_\mu P_\nu^*
         + e \, \bar \psi_\mu \gamma^\nu \gamma^\mu \chi \, P_\nu^*
         + e \, \bar \chi \gamma^\mu \gamma^\nu \psi_\mu \,
                (P_\nu - \bar \psi_\nu \chi ), \zeile
   \Lag2 &=& - e \,  \bar \chi \gamma^\mu D_\mu \chi .
\label{2-Lagrange-new}
\eeql
where we henceforth drop all primes, as the old spin connection
will no longer be used.

For the canonical treatment, we must now perform a space time split
as described at the end of section \rfsec{pure}.
                                                After a little work, one
arrives at the following decomposition of the Lagrangian
\beql
  \Lag0 & = &{}  - \ft12 \eps^{ij} \euo ia \dot \Au ja
        - \eps^{ij} \, \bar \psi_i \dot \psi_j \zeile
  &&{}+ \ft14 (n^{-1} e e^{at} - n^k {e_k}^a) \eps^{ij} F_{ija}
       - \ft12 i\eps^{ij}\,  \bar \psi_i \psi_j \, Q_t
   \zeile    & & {}
  + \ft12 \eps^{ij} ( D_i e_{ja} - \bar \psi_i \gamma_a \psi_j)
    \Ao ta + \eps^{ij} ( D_i \bar \psi_j \psi_t +
                             \bar \psi_t D_i \psi_j ) \zeile
   \Lag1 & = &  {}
 - n \suco P_n \suco P_n^* + n^{-1} \dens h^{ij} \suco P_i \suco P_j^*
  + n^{-1} \eps^{ij} ( \bar \psi_i \, e\gamma^t \chi \, P^*_j
      - \bar \chi \, e \gamma^t \psi_i\,  \suco P_j )
 \zeile     & & {}
 + \eps^{ij} (   \bar \psi_n \gamma_j \chi \,  P_i^*
                -\bar \psi_i \gamma_j \chi \,  P_n^*
                +\bar \chi \gamma_i \psi_n \, \suco P_j
                -\bar \chi \gamma_i \psi_j \, \suco P_n ), \zeile
  \Lag2 & = &{}-  \bar \chi\,  e \gamma^t\,  \dot \chi
             - \ft12 \, \bar \chi\, e\gamma^t \gamma_a \chi\, \Ao ta
             + \ft32 i \, \bar \chi\,e \gamma^t \chi \, Q_t
           \zeile    & & {}
        -  n^{-1} \eps^{ij} \, \bar \chi \, e \g^t \g_i D_j \chi
        -  n^k \, \bar \chi \, e \gamma^t D_k \chi  .
\eeql
which will serve as our starting point for the canonical treatment.

We next compute the canonical momenta, which lead to second class
constraints (to make the formulas less cumbersome, we will not
explicitly indicate the $\x$-dependence below); they are
\beql
   \epuls ai  =  \deltadelta \Lagrange /   \dot \euo ia / = 0,
           \ \  & & \ \
   \Apuls ai  =  \deltadelta \Lagrange /   \dot \Ao ia / =
                     \ft12 \eps^{ij} e_{ja} ,
           \zeile
   \pi^i      = \deltadelta  \Lagrange / \dotbar \psi_i / =  0 ,
          \ \ & & \ \
   \bar \pi^i  = \deltadelta \Lagrange /  \dot \psi_i / =
      - \eps^{ij} \bar \psi_j ,
           \zeile
   \lambda =   \deltadelta \Lagrange / \dotbar \chi / = 0,
         \ \ & & \ \
   \bar\lambda = \deltadelta \Lagrange / \dot \chi / =
         \bar \chi \, e \gamma^t .
\label{2-second-class-momenta}
\eeql
We can now read off the second class constraints
\beql
       P_a^i   :=    \epuls ai, \ \ & & \ \
       Z_a^i   :=    \Apuls ai - \ft12 \eps^{ij} e_{ja} ,\zeile
       \Lambda :=    \lambda,   \ \ & & \ \
   \bar\Lambda :=   \bar \lambda  -   \bar\chi\, e \gamma^t ,\zeile
      \Gamma^i :=   \pi^i, \ \ & & \ \
   \bar\Gamma^i:=   \bar\pi^i +  \eps^{ij} \bar\psi_j .
\label{2-second-class-constraints}
\eeql
As is well known, the Dirac brackets \cite{Dirac} are defined by
\beq
  \dirac A B = \pois A B - \sum_{K,L} \pois A K  C(K,L) \pois L B ,
\label{2-Dirac-formula}
\eeq
where $C(\cdot,\cdot)$ is the inverse of the Poisson matrix defined by
\beq
  \sum_L C(K,L) \pois L M  =  \delta(K,M),
\eeq
Here $K,L,\dots$ label the constraints \rf{2-second-class-constraints}.
A little care has to be taken as we are dealing with fermions here,
because they are anticommuting (Grassmann) variables.
When defining the momenta by \rf{2-second-class-momenta},
$\bar \lambda$ is the {\it negative} momentum  of $\chi$ and the
Hamiltonian reads $ \dotbar \chi \lambda - \bar \lambda \dot \chi
 - \Lagrange$. To get the right equations of motion, the Poisson
brackets, which are symmetric for fermions, must read
$ \pois{ \lambda_\alpha }{\bar \chi_\beta } = - \delta_{\alpha\beta}$,
which in this order corresponds to the bosonic bracket
$ \pois{ \epuls ai } { \euo jb } = - \delta^i_j \delta^b_a $.
Thus the Poisson brackets are always negative if the momentum
is the first entry.

Calculation of the $C(\cdot,\cdot)$ matrix now gives
\beql
  C(Z_a^i,P_b^j) &=& 2 \eps_{ij} \eta^{ab},\zeile
  C(Z_a^i,\Lambda) & = &
     2h^{-1} \eps_{abc} \euo ic \, \bar \chi \gamma^b\ e\gamma^t  ,
     \zeile
  C(\bar\Lambda,\Lambda) & = & - h^{-1} \, e\gamma^t,\zeile
  C(\bar\Gamma^i, \Gamma^j ) & = &  - \eps_{ij}\eins,
\eeql
all other components vanish. As a consistency check, we note
that $C(\cdot,\cdot)$ is antisymmetric if and only if both entries
are bosonic. From these formulas we can deduce the crucial result
that the Dirac bracket between different components of the spin
connection now vanishes, which is not the case for the original
spin connection \cite{Nic91, MN}. This result follows essentially
from the vanishing of $C(Z_a^i, \bar\Lambda)$ and the fact that
$\Lambda$ does not depend on the dreibein. Furthermore, it is also
easy to check that the spin connection now commutes with $\chi$.
However, it does {\it not} commute with $\bar \chi$, which is
therefore not a good canonical variable. For this reason, we will
not use $\bar \chi$, but rather $\bar \l$ as an independent phase
space variable; the two fields are related by $\bar \l = e
\bar \chi \g^t$ by \rf{2-second-class-constraints}.
All non-vanishing Dirac brackets
are then numerical and given by
\beql
  \dirac{ \Ao ia (\x) }
        { \euo jb(\y) }  & = &
             2 \eps_{ij} \eta^{ab}\,  \deltaxy  \zeile
  \dirac{  \chi_\alpha(\x)  } {  \bar\lambda_\beta(\y) }
     & = &-\delta_{\alpha\beta}     \, \deltaxy \zeile
  \dirac{  \psi_{i\alpha}(\x) } { \bar \psi_{j\beta}(\y) }
     & = &   \eps_{ij} \delta_{\alpha\beta} \, \deltaxy
\label{2-Dirac-brackets}
\eeql
We repeat that the absence of the complex conjugate connection
$({A_\m}^a)^*$ from the constraints is an important consistency check
on our results, since this field would have non-vanishing brackets
with both ${A_\m}^a$ and $\chi$.

Next we proceed to the discussion of the first class constraints.
Defining the momenta of the scalar fields as in \rf{SIGMAImpuls2}, but
with the complex notation introduced above, we have
\beql
    P  &=& \deltadelta \Lagrange / P_t^* /  =
    {}      - n P_n  + n\, \bar \psi_n \chi
          - \eps^{ij} \, \bar \psi_i \gamma_j \chi ,
       \zeile
    P^*&=& \deltadelta \Lagrange / P_t   /  =
    {} -   n P^*_n + n\,  \bar \chi \psi_n
          - \eps^{ij}\,  \bar \chi \gamma_i \psi_j ,
       \zeile
    Q  &=& \deltadelta \Lagrange / Q_t   /  =
    {}  - \ft12 i \eps^{ij} \, \bar \psi_i \psi_j
        + \ft32 i \, \bar \chi \, e \gamma^t \chi  .
\label{2-V-momenta}
\eeql
{}From \rf{LIE-Poisson-matrix} we obtain their Poisson (or Dirac)
brackets
\beqx
   \pois { \V } { P    } =  \V Z    , \ \ \ \
   \pois { \V } { P^*  } =  \V Z^*  , \ \ \ \
   \pois { \V } { Q    } =  \V Y    ,
\eeqx
\beq
   \pois { P  } { P^*  } =     2 i Q  , \ \ \ \
   \pois { Q  } { P    } =  - 2 i P , \ \ \ \
   \pois { Q  } { P^*  } =     2 i P^* .
\label{2-V-brackets}
\eeq
These brackets together with \rf{2-Dirac-brackets} constitute the
complete list of non-vanishing Dirac brackets of $N=2$ supergravity
(brackets that have not been listed vanish). To summarize,
our basic canonical variables are
\beq
  \euo ia , \ \Ao ia ,\ \V , \  P, \ P^*, \ Q , \
  \bar\lambda , \ \chi , \ \bar\psi_i , \ \psi_i,
\label{2-canonical variables}
\eeq
Observe that quantization is now straightforward to implement by
replacing the momenta by functional differential operators.
The Lagrange multipliers leading to first class constraints are
\beq
  n^{-1} , \ n^i , \ \Au ta , \ \bar \psi_t , \ \psi_t.
\label{2-Lagrange-multipliers}
\eeq
%For later use we compute the brackets of the momenta with the spatial
%derivatives. This yields
%\beql
%  \pois{ P_i^* \vonx } { P   \vony } &=&
%    \partial_i \deltaxy + 2i Q_i \, \deltaxy \zeile
%  \pois{ Q_i  \vonx  } { P   \vony } & = &
%    2 i P_i \, \deltaxy                      \zeile
%  \pois{ P_i   \vonx } { P^* \vony } & = &
%    \partial_i \deltaxy - 2i Q_i \, \deltaxy \zeile
%  \pois{ Q_i  \vonx  } { P^* \vony } & = &
%   -2 i P_i^* \, \deltaxy                   \zeile
%  \pois{ P_i   \vonx } { Q   \vony } & = &
%    2 i P_i \, \deltaxy                     \zeile
%  \pois{ P_i^* \vonx } { Q   \vony } & = &
%    - 2i P_i^* \, \deltaxy                  \zeile
%  \pois{ Q_i  \vonx  } { Q   \vony }  & = &
%   \partial_i  \, \deltaxy
%\eeql

As the momentum $Q$ does not contain any time derivative in
\rf{2-V-momenta}, we have the first class constraint
\beq
  \T  = {}- Q - \ft12 i \eps^{ij} \, \bar \psi_i \psi_j
             + \ft32 i \, \bar \lambda \chi ,
\label{2-U(1)-constraint}
\eeq
which is the generator of local $\grp U(1)$ transformations.
The other first class constraints are obtained by varying
the Lagrangian with respect to the multipliers
\rf{2-Lagrange-multipliers}. $\Ao ta $ yields the
Lorentz constraint
\beq
  \L_a = \ft12 \eps^{ij} D_i \euu ja
    - \ft12 \eps^{ij} \, \bar \psi_i \gamma_a \psi_j
    - \ft12 \bar \lambda \gamma_a \chi .
\label{2-Lorentz-constraint}
\eeq
The multipliers $\bar \psi_t $ and $\psi_t$ correspond to
the supersymmetry constraints
\beql
   \S &=& \eps^{ij} D_i \psi_j
        - \chi \, P^*  - \eps^{ij} \, \gamma_i \chi \, P_j^*,\zeile
   \bar \S & = &
          \eps^{ij} D_i \bar \psi_j
        - P \, \bar \chi + \eps^{ij} P_j \, \bar \chi \gamma_i
        + \eps^{ij} \, \bar\psi_i
          (\chi \, \bar \chi \gamma_j - \gamma_j \chi \, \bar\chi).
\label{2-susy-constraints}
\eeql
The asymmetry in these expressions is caused by the fact that it is
the redefined connection $\Ao ia$ which appears in the derivative on
$\bar \psi_j$ in the second line (and not ${\Ao ia}^*$!), but
in fact, the second expression is the Dirac conjugate of the first
because of the Fierz identity
\beq
\eps^{ij} \left( \bar \psi_i \chi \, \bar \chi \g_j -
        \bar \psi_i \g_j \chi \, \bar \chi \right) =
 - \eps^{abc} \eps^{ij} e_{jb} \bar \chi \g_c \chi  \,
    \bar \psi_j \g_a
\eeq
The first constraint in \rf{2-susy-constraints}
is manifestly polynomial; to render the second
polynomial as well, we multiply it by $e \g^t$ from the right so
as to replace $\bar \chi$ by $\bar \l$. In this way, we get
\beq
   \dens \S := \bar \S \, e \gamma^t =
          \eps^{ij} \, D_i \bar \psi_j  \, e\gamma^t
        - P \, \bar \lambda - \eps^{ij} P_j \, \bar \lambda \gamma_i
        - \eps^{ij} \, \bar\psi_i
          (\chi \, \bar\lambda\gamma_j + \gamma_j \chi \,\bar\lambda).
\label{2-dens-susy-constraint}
\eeq

The derivative of the Lagrangian with respect to $n^k$ and
$n^{-1}$ gives the diffeomorphism and Hamiltonian constraints:
\beql
  \H_k^\prime &=& - \ft14  \euo ka    \eps^{ij} F_{ija}
             -   \bar \lambda   D_k \chi
             + \suco P \suco P^*_k   +  \suco P_k \suco P^* \zeile
        & & {} +   \eps^{ij} P_i^* \bar\psi_j \gamma_k \chi
               -   \eps^{ij} \suco P_i \bar \chi \gamma_k \psi_j,
             \zeile
    \H^\prime &=&  \ft14 e\eoo ta \,  \eps^{ij} F_{ija}
         -  \eps^{ij} \, \bar \lambda \gamma_i D_j \chi
          + \suco P \suco P^* + \dens h^{ij} \suco P_i \suco P_j^*
                                               \zeile
  & & {} - \eps^{ij} P_i^* \, \bar\psi_j e \gamma^t \chi
         + \eps^{ij} \suco P_i \, \bar \lambda   \psi_j  ,
\label{2-Hamilton-constraints-prime}
\eeql
where we used the supercovariant quantities
\beqx
    \suco P_i = P_i - \bar \psi_i \chi, \ \ \ \
    \suco P   = P + \eps^{ij} \, \bar\psi_i \gamma_j \chi ,
\eeqx
\beq
    \suco P_i^* = P_i^* - \bar \chi \psi_i , \ \ \ \
    \suco P^*   = P^* + \eps^{ij} \, \bar\chi  \gamma_i \psi_j .
\label{2-P-supercovariant}
\eeq
The covariant momentum $\suco P$ is nothing but the time
component of the quantity already defined in \rf{2-susy-trafo},
as can be seen by \rf{2-V-momenta}.

At first sight it seems rather difficult to cast these constraints
into a polynomial form, as
$\bar \chi $ appears also implicitly through $\suco P^*$ and
$\suco P^*_i$, but indeed all terms containing $\bar \chi$ can be
eliminated by adding suitable multiples of other constraints.
               Let us first consider the diffeomorphism constraint.
The terms containing $\bar \chi$ are
\beq
        - \suco P \, \bar \chi \psi_k
  - \eps^{ij} \suco P_i \, \bar \chi \gamma_k \psi_j
  + \eps^{ij} \suco P_k \, \bar \chi \gamma_i \psi_j.
\eeq
A short calculation shows that this is equal to
\beq
  \bar \S \psi_k - \eps^{ij} \, D_i \bar \psi_j \psi_k .
\eeq
Thus if we subtract $\bar \S \psi_k$ from $\H_k^\prime$, all
the $\bar\chi$ terms disappear and we get the polynomial constraint
\beql
  \H_k  & =& - \ft14 \euo ka  \eps^{ij} F_{ija}
             - \bar \lambda  D_k \chi
             - \eps^{ij} \, D_i \bar \psi_j \psi_k  \zeile
      & & {} +  P^*_k \suco P +  P^* \suco P_k
             + \eps^{ij} P_i^* \, \bar\psi_j \gamma_k \chi
\label{2-diffeomorphism-constraint}
\eeql
As usual this is not the real generator of spatial diffeomorphisms,
but it generates extra $\grp U(1)$, Lorentz, and supersymmetry
transformations. The generator of pure translations is
\beq
   \D_k =  \H_k - Q_k \T - \Ao ka \L_a - \bar \psi_k \S,
\eeq
and reads explicitly
\beql
   \D_k & = &
   -  \ft12 \eps^{ij}
      (\partial_i \Ao ja \euu ka +  \Ao ka  \partial_i \euu ja )
   -  \eps^{ij}
      (\partial_i \bar \psi_j \psi_k + \bar \psi_k \partial_i \psi_j)
                                          \zeile    && {}
            +  P P_k^* + P^* P_k + Q Q_k
   -   \bar \lambda \partial_k \chi  .
\label{2-real-diffeomorphism-constraint}
\eeql

The situation is similar for the Hamiltonian constraint.
The $\bar \chi$ terms to be subtracted are
\beq
   \eps^{ij} \suco P \, \bar \chi \gamma_i \psi_j
   - \dens h^{ij} \suco P_i \, \bar \chi \psi_j .
\eeq
Again this nonpolynomial expression can be rendered polynomial
by subtracting a suitable multiple of $\bar \S$. It is
equal to
\beq
  - \eps^{ij} \, \bar \S \gamma_i \psi_j +
  \eps^{ij}\eps^{kl} \, D_k \bar\psi_l \gamma_i \psi_j
  +  \eps^{ij} \suco P_i \, \bar \lambda  \psi_j .
\eeq
Inserting this we get the Hamiltonian constraint
\beql
    \H & =&  \ft14   e \eoo ta \, \eps^{ij} F_{ija}
             + \eps^{ij} \,  \bar \lambda \gamma_j D_i \chi
             + P^*  \suco P
             + \dens h^{ij} P_i^* \suco P_j      \zeile
    & & {}   + 2 \eps^{ij} \suco P_i \, \bar \lambda \psi_j
             + \eps^{ij} \eps^{kl} \,
                 D_i \bar \psi_j \gamma_k \psi_l
              - \eps^{ij} P_i^* \, \bar\psi_j\, e\gamma^t \chi .
\label{2-Hamilton-constraint}
\eeql

We can now compute the classical Dirac bracket algebra of constraints
and verify that it closes. The brackets between the ``kinematical"
constraints $\L$, $T$, $\H_k$ and $\S$ are straightforward and
yield the expected results. The brackets between the supersymmetry
generators require more work, and, in particular, repeated use of the
Fierz identities quoted in section 2. After a somewhat lengthy
calculation, one obtains
\beql
  \dirac{  \S[\bareps] }{ \dens \S[\eta] } & = &
     \int \d^2 \x \, \big (
       - \bareps \eta \, \H  + \eps^{kl}  \H_k  \,
       \bareps \gamma_l \eta
       + 2 i \bar \lambda \eta \, \bareps \chi \, \T
       + 2 \bar \lambda \eta \, \bareps \gamma^a \chi \, \L_a \big),
  \zeile
  \dirac{  \S[\bareps] }{ \S[\bareps^\prime] } & = & 0,  \zeile
  \dirac{  \dens\S[\eta] }{ \dens\S[\eta^\prime]}& = &
   2 \int \d^2 \x \,  \eps^{ij} \euo ia \big(
          \bar \psi_j \eta \, \dens \S \gamma_a \eta^\prime
        + \bar \psi_j \gamma_a \eta \, \dens \S \eta^\prime \big).
\label{2-super-algebra}
\eeql
where, for convenience, the supersymmetry generators have been smeared
with smooth spinorial (i.e. anti-commuting) test functions
$\bareps (\x )$ and $\eta (\x )$ according to
\beq
\S [\bareps ] := \int \d^2 \x \, \bareps (\x ) \S (\x )  \;\;\;\; ,
\;\;\;\; \dens \S [\eta ] := \int \d^2 \x \, \dens \S (\x ) \eta (\x )
\eeq
The formulas \rf{2-super-algebra} show that indeed all the bosonic
constraints can be generated from the fermionic ones, and in this
sense, the supersymmetry generators can be thought of as the square
roots of the bosonic ones. We note that a complete check of closure
would also require the determination of the brackets involving the
Hamiltonian constraint, but we omit this consistency check.

Finally, one can verify that the constraints are the canonical
generators of the associated space dependent gauge transformations
on the fields as expected. Since this computation is completely
straightforward, we refrain from giving further details.

\Section{Quantization and Constraint Algebra}
\labelsec{quant}

We will now perform the canonical quantization and
re-examine the relations \rf{2-super-algebra} in the context of the
quantum theory.
After expressing all the first class constraints as polynomials
in terms of the canonical variables \rf{2-canonical variables},
it is straightforward to give a quantum operator
representation for them.
We use the convention that the commutator is obtained by multiplying
the Dirac bracket by $-i$. We will also define the quantum constraints
to be $i$ times the classical constraints. In this way they generate
the same transformations on the fields
as the classical constraints and no extra factors will appear in
the constraint algebra.

For each pair of canonically conjugate variables, we have to
choose one which is to be represented by a multiplication operator.
The bosonic fields and their momenta $P$, $P^*$ and $Q$
will be represented by the matrix $\V$ and the differential operators
\rf{LIE-quantum-operator}, respectively. In this representation,
the wave functional depends on the scalars via the matrix $\V (\x)$.
As we explained in section 3, we could equivalently use
the coordinates $(\ve^m, u)$ on $\grp SL(2,\R)$ (where $m=1,2$, and $u$
parametrizes the subgroup $\grp SO(2)$) as
multiplication operators and represent the momenta by
\rf{OPDarstellung2}.
For the remaining operators we are in principle free to choose
$\euo ia$ or $\Ao ia$, $\bar \lambda$ or $\chi$, $\bar \psi_i$
or $\psi_i$ as multiplication operators; of course, it is very
likely that the resulting quantum theories will be inequivalent,
depending on this choice. We will here adopt an operator representation
which renders all the constraints homogeneous in the functional
differential operators (see below), and which is given by
\beql
      \euo ia &\longrightarrow&  -2i\eps_{ij} \deltadelta / \Au ja /,
             \zeile
      \bar\lambda_\alpha & \longrightarrow &
        i \deltadelta / \chi_\alpha / ,
             \zeile
      \psi_{i\alpha} & \longrightarrow &
        - i \eps_{ij} \deltadelta / \bar \psi_{j\alpha} /.
\label{Q-operator-rep1}
\eeql
The wave functional $\Psi (\phi )$ will thus depend on the fields
$\Ao ia , \V , \chi , \bar \psi_i$, which we will collectively denote
by $\phi$. With the above operator representation the Lorentz,
$\grp U(1)$ and diffeomorphism constraints become well defined
if ordered in such a way that all differential operators appear
{\it to the right}. In particular, they will then generate the
respective space dependent gauge transformations on the fields
(i.e. without extra anomalous contributions). Their algebra reads
\beql
  \komm{ \D[n^k] }{\D[m^k] } &=&
         \D[ m^l \partial_l n^k -  n^l \partial_l m^k ],\zeile
  \komm{ \L[\omega^a] }{\L[\upsilon^a] } &=&
         \L[\eps^{abc} \omega_b \upsilon_c ],  \zeile
  \komm{ \D[n^k] }{\L[\omega^a] }  &=&
         \L[-n^l \partial_l \omega^a  ],  \zeile
  \komm{ \D[\omega^a] }{\T[q] } &=&
         \T[-n^l \partial_l q ].
\label{Q-subalgebra}
\eeql
where $\D[n^k] := \int \d^2 \x \, n^k (\x ) \D_k (\x )$, etc.
Remember that the quantum constraints are
defined as $i$ times the classical constraints, thus there are
no extra factors of $i$ in the algebra.

Next, we turn to the supersymmetry generators $\S$ and $\dens \S$.
In the representation \rf{Q-operator-rep1}, $\S$ becomes a first
order differential operator; smearing with an anticommuting
spinor paramater $\bareps$ as in (4.39), we get
\beq
   \S[\bareps]  = \int \d^2 \x \left(
          D_i \bareps \deltadelta / \bar\psi_i /
      + \bareps \chi \, \Tr \left( \V Y^* \deltadelta / \V / \right)
 - 2 \bareps \gamma^a \chi \, P_i \, \deltadelta / \Ao ia / \right) .
\label{Q-super-rep}
\eeq
{}From this expression, it is easy to see that there are no ordering
ambiguities and no short distance singularities in $\S$.
In this form $\S$ can be regarded as a ``kinematical" constraint
whose action on the wave functional is just given by the
transformation of the fields under supersymmetry. Below we will see
that it is even possible to exponentiate this generator to obtain
a finite supersymmetry transformation.

The constraint $\dens \S$, on the other hand, is a second order
operator in our representation and is therefore ``dynamical" like
the Hamiltonian constraint $\H$. Explicitly, we have
\beql
    \dens\S[\eta] & = &  \int \d^2 \x \bigg(
      4 i \eps^{abc} \, D_i \bar\psi_j  \gamma_a \eta \,
        \deltadelta / \Ao ib /
     \, \deltadelta / \Ao jc /
 \ + \  i \eta_\alpha \, \Tr \left( \V Y \deltadelta /\V/ \right)
      \, \deltadelta / \chi_\alpha /  \zeile
      & & {} - 2 i \Big(
          \eta_\alpha \, (\bar \psi_i \gamma^a \chi)
        - (\gamma^a \eta)_\alpha  \, (P_i - \bar\psi_i \chi) \Big)
      \, \deltadelta / \Ao ia /
      \, \deltadelta / \chi_\alpha / \bigg) .
\label{Q-dens-super-rep}
\eeql
In contrast to $\S$, $\dens \S$ must be regularized because it contains
products of functional differential operators at coincident points.
This can be done for instance by smearing all operators with
a regularized delta function $\delta_\rpar (\x ,\y )$ according to
\beq
{\cal O} (\x ) \longrightarrow {\cal O}_\rpar (\x) :=
\int \d^2 \y \, \delta_\rpar (\x , \y ) {\cal O} (\y )    ,
\eeq
where $\rpar$ is a regularization parameter such that $\lim_{\rpar
\rightarrow 0} \delta_\rpar (\x ,\y ) = \delta (\x , \y )$. Denoting
the regularized constraint by $\dens\S_\rpar$, we automatically
obtain a regularized Hamiltonian from the regularized commutator
of $\S$ with $\dens\S_\rpar$ that is obtained from
\rf{2-super-algebra}. In the remainder, we will simply assume
that our formal manipulations can be made rigorous by means of
a suitable regularization.

In addition to the problem of regularization there is also
an operator ordering ambiguity in the definition of
$\dens \S$, as one can place the differential operators either to the
left or to the right \footnote{Or in between.}. Whichever prescription
we choose, we then {\it define} the ordering of the
bosonic constraint operators by the right hand side of the commutator
\rf{2-super-algebra}, so all ambiguities disappear once we have fixed
the ordering of $\dens \S$. Relying on formal manipulations,
one can convince oneself that any change of ordering in
$\dens \S$ produces a singular term proportional to
\beq
    i \delta(\x,\x)  \, \bar\psi_i \gamma^a \eta \,
        \deltadelta / \Ao ia /
\label{Q-ordering-extra}
\eeq
where the overall factor depends on which terms are interchanged.
In \rf{Q-dens-super-rep}, all differential operators have been
placed to the right. Remarkably, and in contrast to
ordinary gravity, it turns out that the singular contributions
precisely cancel when one inverts this
ordering by placing all differential operators to the left,
so that the two ordering prescriptions in fact coincide!
This change of ordering also reverses the order of canonically
conjugate operators on the right hand side of the commutators in
\rf{2-super-algebra}.

In order to be able to interpret the quantum constraints
$\S \Psi = \dens \S \Psi = 0$ as square roots of the quantized bosonic
constraints, and especially the Wheeler DeWitt equation
$\H \Psi =0$, we would like to find an operator ordering prescription
such that in the quantized version of \rf{2-super-algebra}, the
constraint operators on the right hand side always appear {\it to the
right} of the field dependent structure functions. Otherwise, there
will be extra contributions from the bosonic constraint operators
acting on the structure functions, when \rf{2-super-algebra} acts
on a wave functional, and the constraint algebra will be ``anomalous".
Let us thus put all differential operators to the right as in
\rf{Q-dens-super-rep} and recalculate the commutator
\rf{2-super-algebra}, now paying attention to the order in which
the operators appear. After some algebra, we arrive at the same result
with exactly the ordering indicated in \rf{2-super-algebra}.
In particular, the operator $T$ is properly ordered, i.e. with
the differential operators to the right; this is important because
otherwise $T$ would not only generate $\grp U(1)$ transformations
on the fields but extra singular terms (the ordering does not matter
for the Lorentz generator $L_a$ because $\g_a$ is traceless). We also
observe that the last commutator in \rf{2-super-algebra} remains
the same if $\dens \S$ is placed to the left because the singular
terms again cancel by virtue of the Fierz identity
\beq
\eps^{ij} \bar \psi_i \g^a \eta \, \bar \psi_j \g_a \eta' =
\eps^{ij} \bar \psi_i \eta \, \bar \psi_j \eta'
\eeq

In summary, all constraint operators appear in the desired order,
except for the diffeomorphism generator, which appears in the ``wrong"
order, i.e. to the left of a structure function.
Consequently, the equations $\S \Psi = \dens \S \Psi = 0$ imply
that $\H \Psi = L \Psi = T \Psi = 0$, but {\it not} $\H_k \Psi =0$.
This means that a solution of the supersymmetry constraints cannot be
diffeomorphism invariant, although it would satisfy the other
constraints. We note that a similar ``anomaly" was already encountered
in \cite{JacSmol}, and was there identified as the basic reason
why the solutions of \cite{JacSmol} fail to be diffeomorphism
invariant, despite the fact that they are formally annihilated
by the Hamiltonian constraint. Another somewhat bothersome
feature is that the ordering inside
$\H_k$ as obtained from this commutator is precisely as in
\rf{2-real-diffeomorphism-constraint}, i.e. with $P^*$ to the
left of $P_k$ (all the other operators in $\H_k$ are in the
``correct" order). This means that in addition to diffeomorphisms
$\H_k$ (or rather $\D_k$) will generate anomalous singular terms
when acting on the scalar fields.

We could now try to cancel the anomalous contribution
by some intermediate reordering of the constraint operators
(note that simply reverting the order of the operators
will not do, because both $\S$ and $\dens \S$ remain the same as we
already explained). Commuting $\H_k$ through
$\g_l = \euo l a  \g_a$ to the right, we pick up a singular term
\beq
\e^{kl} \H_k \, \bar \e \g_l \eta = \e^{kl} \bar \e \g_l \eta \, \H_k
+ \left( \bar \e \g^a \eta \, \bar \l \g_a \chi - 2\bar \e \g^a \eta
\, L_a \right) \delta (\x , \x )
\eeq
where we disregard a term involving a derivative on $\delta (\x,\x)$.
Here the third term can be ignored, as it is proportional to the
Lorentz constraint $L_a$, but the second term constitutes an anomaly.
It can be cancelled by a suitable reordering ``inside" the constraint
operators. Unfortunately, apart from the rather artificial orderings
required for this cancellation to work, this procedure leads to new
``anomalies", which must be cancelled in turn. We have so far not
found any ordering prescription that would remove {\it all} anomalies
and maintain the desired ordering of the quantum constraint algebra,
but we anyhow would not expect that this problem can be solved simply
by a clever reordering of the operators. What is really needed, but
unfortunately unavailable at this point, is a properly defined scalar
product on the space of physical states.

Given this state of affairs, we believe that a more
reasonable option is therefore to give up diffeomorphism invariance (at
least in any conventional sense) of the wave functional, replacing the
the constraints $\H$ and $\H_k$ by a single (matrix valued)
new constraint $\K:= \H - \eps^{kl} \H_k \gamma_l$, whose
ordering is defined by the commutator of $\S$ with $\dens\S$, but
whose physical significance is obscure. In the remainder, we
will tentatively adopt this point of view, and
discuss possible ans\"atze to solve the supersymmetry constraints.
We have already seen that $\S$ is
homogeneous of the first degree in the functional differential
operators, while the conjugate constraint operator
$\dens\S$ is homogeneous of the second degree, and therefore
much harder to solve.
We will now demonstrate how the constraint $\S \Psi [\phi ] =0 $
can be solved in full generality by the functional analogue of the
method of characteristics known from the theory of first order
partial differential equations \cite{CH}. Before going into the
details, however, two remarks are in order. First of all,
$\Psi \equiv 1$  trivially solves all the constraints if the
differential operators are placed to the right (this is not true
if another and inequivalent ordering is chosen).
Secondly, given one non-trivial solution (i.e. $\Psi \neq 1$), we can
construct many more solutions by application of the conserved charge
$\cal Q$ constructed at the end of section 3 (cf. \rf{chargerep}).
By use of \rf{chargeconstraint}, one shows that
\beq
\Big[ {\cal Q} \, , \, \S [\bareps ] \Big] =
\Big[ {\cal Q} \, , \, \dens \S [\eta] \Big] = 0
\eeq
Therefore ${\cal Q} \Psi$ solves the supersymmetry constraints
if $\Psi$ does. It is in this sense that $\grp SL(2,\R )$ (and
the corresponding hidden symmetry groups for higher $N$ supergravities)
can be viewed as ``solution generating symmetries" of the quantum
constraints \cite{Nic91}. It would be interesting to check whether
this symmetry is unitarily realized on the space of physical states,
in which case one could even generate infinitely many solutions out
of a given non-trivial one. However, this question again hinges on
the unresolved problem of the scalar product.

In order to use the method of characteristics, we must determine
the orbits in functional space generated by the action of the
operator $\S$. This is equivalent to exponentiating an infinitesimal
local supersymmetry transformation with parameter $\bx =\bx (\x )$
so as to obtain the corresponding {\it finite} local supersymmetry
transformation. Although this would be a formidable problem in general,
in the case at hand the solutions can be obtained in closed form,
because repeated application of the supersymmetry generator $\S$ on
any of the fields gives zero after at most three steps.
Labeling the ``initial" fields by the superscript
$^{(0)}$, we thus find
\beql
\chi (\bx ) &=& \chi^{(0)}, \zeile
\V (\bx ) &=& \V^{(0)} \big(1 + \bx \chi^{(0)} Y^* \big)  \zeile
&\Rightarrow& P_i^* (\bx ) = P_i^{*(0)}     \zeile
&&\suco P_i (\bx ) = \suco P_i^{(0)} + \bx D_i^{(0)} \chi^{(0)} \zeile
&&Q_i(\bx ) =  Q_i^{(0)} - 2iP_i^{*(0)} \, \bx \chi^{(0)}  \zeile
{A_i}^a (\bx )  &=& A_i^{(0) a}
                    - 2 P_i^{*(0)} \bx \g^a \chi^{(0)}  \zeile
\bar \psi_i (\bx ) &=& D_i^{(0)} \bx - 2P_i^{*(0)} \, (\bx \chi^{(0)})
      \bx + \bar \psi_i^{(0)}
\label{SUSY-Trajektorien}
\eeql
Hence $P_i^*$ and $\chi$ are inert, whereas the other fields transform
in a relatively simple fashion\footnote{Here we assume that all
the fields are complexified to make the {\it chiral} supersymmetry
transformation well defined} since only the gravitino field
evolves with terms quadratic in $\bx$. The trajectories described by
the fields $\phi(\bx)$ in functional space as $\bx$ is varied
are the orbits under local supersymmetry. This fact is reflected
in the identity
\beq
 \S [\bar \e , \phi (\bx )] \Psi \big[ \phi (\bx ) \big]  =
 \int \d^2 \x \, \bar \e (\x ) {\delta \over {\delta \bx (\x )}}
\Psi \big[ \phi (\bx ) \big]
\eeq
The most general solution of the quantum constraint $ \S \Psi =0$
is now obtained by choosing $\Psi$ such that ${\delta \over {\delta
\bx}} \Psi [ \phi (\bx )] = 0$, i.e. constant along the trajectories
given by the ``evolution equations" \rf{SUSY-Trajektorien},
and by prescribing arbitrary values of $\Psi$ on some
(infinite dimensional) hypersurface in functional space which
is nowhere tangent to the supersymmetry orbits.

In practice, choosing a functional hypersurface amounts to
imposing some gauge condition on the gravitino. For example,
let us choose $c^i \bar \psi_i = 0$, where $c^i (\x)$ is an
arbitrary non-vanishing vector field. Given an arbitrary
configuration of the fields $\phi$ (which, in general, will not
satisfy the gauge condition), we must first determine to which
supersymmetry orbit it belongs. This requires solving the
first order partial differential equation for $\bx$
\beq
c^i \bar \psi_i (\bx) = c^i \Big( D_i \bx - 2P_i^* (\bx \chi ) \bx
 + \bar \psi_i \Big) = 0
\label{gauge-fix}
\eeq
We note the following subtleties about this (and similar) gauge
conditions. For arbitrary and topologically non-trivial space-like
surfaces, the vector field $c^i (\x )$ will in general have zeros,
at which the gauge condition degenerates. Secondly, a given
$\bar \psi_i$ is gauge equivalent to a configuration satisfying
$c^i \bar \psi_i = 0$ if and only if the solution $\bx$ is single
valued on the spacelike surface, i.e. obeys
$\oint_\g dx^i \partial_i \bx = 0$ for any closed curve $\g$.
Then $\bx$, which depends on the initial fields $\phi$, is the
finite supersymmetry transformation parameter connecting the given
configuration of fields to the gauge hypersurface. We now simply define
\beq
\Psi \big[ \phi \big] :=
\Psi \big[ \phi (\bx) \big]
\eeq
with $\bx$ from \rf{gauge-fix}, where
$\Psi [\phi (\bx )]$ is given by the previously assigned value
of $\Psi$ on the gauge hypersurface. Incidentally, the differential
equation \rf{gauge-fix} is again solved by the method
of characteristics, but now in ordinary, not in functional space.
Consequently, the transformation parameter $\bx$ is completely
determined by \rf{gauge-fix} only after specification of suitable
initial values.

To gain a somewhat different perspective (and possibly
also to establish a connection with the work
of \cite{JacSmol, RovSmol}), we will now consider wave
functionals that do not depend on the fields on the whole spatial
surface, but are supported only on given set of curves $\k(s)$
determined from the equations $\dot x^i (s) = c^i (\x (s))$
where $c^i (\x )$ is the vector field introduced in \rf{gauge-fix}
(the integral curves $x^i (s)$ are just the characteristics of
\rf{gauge-fix}). Along $\k(s)$, we now consider the
${\grp SL(2,\R )} \times \grp SO(2)$ valued gauge potential
(remember $\grp SO(2)$ is just the helicity group of the $N=1$
supergravity in four dimensions)
\beq
{\cal A}_i := \ft12 \g_a \left( {A_i}^a
                 - 2P_i^* \bx \g^a \chi \right)
  + \ft12 i \left( Q_i - 2i P_i^* \bx \chi  \right)
\eeq
where $\bx = \bx (\phi )$ is determined from \rf{gauge-fix}. Note
that the connection ${\cal A}_i$ also depends on the bosonic and
fermionic matter fields; it is just the
gauge potential occurring in the covariant derivative on
$\bar \psi_i$ shifted by a finite supersymmetry transformation
according to \rf{SUSY-Trajektorien}. Next introduce the path
ordered integral
\beq
T_\k ({\bf a},{\bf b}) = \Pexp \int_0^1 ds \, c^i (\x (s))
            {\cal A}_i (\x (s))
\eeq
where the curve $\k$ is parametrized such that $\k (0) = {\bf a}$
and $\k (1) = {\bf b}$ are the initial and end points
connected by the curve $\k (s)$, respectively.
Computing the variation of $T_\k$ under local supersymmetry, we obtain
\beq
\Big[ \S [\bareps ] \, , \, T_\k ({\bf a}, {\bf b} ) \Big] =
\int_0^1 ds \, c^i (\x (s)) T_\k ({\bf a}, \x (s)) \, \delta
      {\cal A}_i (\x (s)) \, T_k (\x (s), {\bf b})
\label{vary-Wilson}
\eeq
with
\beq
\delta {\cal A}_i  = - \g_a P_i^* \bareps \g^a \chi +
                         P_i^* \bareps \chi
+P_i^* \left( \delta \bx \chi - \delta \bx \g^a \chi \, \g_a \right)
\eeq
Here, the variation $\delta \bx$ must be determined from equation
\rf{gauge-fix}. Namely, varying \rf{gauge-fix}, we find that after a
little algebra and use of the identity $2\bx \chi \, \bareps =
-\bareps \chi \, \bx - \bareps \g^a \chi \, \bx \g_a$, this
equation reduces to
\beq
c^i D_i \big( {\cal A}_i (\bx )\big) \Big( \delta \bx
   + \bareps \Big) = 0
\eeq
The solution is thus
\beq
\delta \bx (\x ) = - \bareps (\x ) +
\big( \delta \bx + \bareps \big) ({\bf a}) T_\k ({\bf a}, \x )
\eeq
where $\x = \x (s)$ is any point along the curve. If the initial
value $\bx_0 \equiv \bx ({\bf a})$ is chosen in such a way
that $\delta \bx_0 + \bareps ({\bf a}) = 0$, the gauge potential
${\cal A}_i$ is invariant, and the right hand side of \rf{vary-Wilson}
vanishes. A solution, which is also invariant under the $U(1)$ and
Lorentz constraints is then easily obtained by closing the curve
into a loop and taking the trace. These solutions can be regarded as
``supercovariant" extensions of the solutions found in \cite{JacSmol}.
We have not checked, however, whether they
are also annihilated by the second order constraint $\dens \S$.
We note that for $\delta \bx (\phi) = - \bareps$, one can construct many
other invariants: since all the $\phi (\bx )$ in \rf{SUSY-Trajektorien}
are invariant with this choice of $\bx$, an arbitrary functional
of them will also be invariant. In this respect, there is nothing
special about loop functionals as opposed to functionals that are
supported on the whole space-like surface.

\Appendix A {Decoupling of Canonical Variables for $N>2$}

In this appendix we explain how the redefinition of the spin
connection works for the higher $N$ theories. We will not
discuss these models in detail here, but refer the reader
to \cite{dWNT} for further explanations. In analogy with
\rf{2-new-connection}, we proceed from the ansatz
\beq
    \AAo \mu a := \Ao \mu a + \Bo \mu a , \ \ \ \
    \Bo \mu a := \ft12  i \eps^{abc} \euu \mu b \,
          \bch A \gamma_c \ch B  \, \JAB ,
\label{N-new-connection}
\eeq
where $\ch A$ are the matter fermions, which transform as spinors
under $H$. Observe that the new spin connection will be {\it complex}
just as for $N=2$. $\J_{\dot A \dot B}$ is an antisymmetric matrix to be
determined by requiring decoupling of the canonical variables;
we will find that $\JAB$ is a complex
structure, i.e. $\J^2 = -{\bf 1}$. Inserting this ansatz into
\rf{3Feldstaerke}, we obtain
\beq
  F_{\mu\nu a} = F^\neu_{\mu\nu a}
     - 2 \, D^\neu_{[\mu} \Bo {\nu]}a
     - \eps_{abc} \Bo \mu b \Bo \nu c.
\eeq
where the derivative $D'_\m$ is covariant with respect
to the Lorentz group $\grp SO(2,1)$
and the gauge group $H$ (see \cite{dWNT} for the precise definitions).
As in section \rfsec{N=2}, we may replace the original spin connection
by $\AAo \mu a$ in the Lagrangian.
Up to a total derivative we then have
\beql
  \ft14 \eps^{\mu\nu\rho} \euo \mu a F_{\nu\rho a} & = &
  \ft14 \eps^{\mu\nu\rho} \euo \mu a F_{\nu\rho a}^\neu
- \ft14 i e \eou \mu a \,
             D_\mu^\neu (\, \bch A \gamma^a \ch B  \, \JAB )
  \zeile &&{}   \ \ \ \ \ \ \ \ \ \ \
  - \ft14 e \, \bch A   \ch B \, \J_{\dot B \dot C} \,
        \bch C \ch D \, \J_{\dot D \dot A},\zeile
  \ft12 \eps^{\mu\nu\rho} \, \bar \psi_\mu^I D_\nu \psi_\rho^I & = &
  \ft12 \eps^{\mu\nu\rho} \, \bar \psi_\mu^I D_\nu^\neu \psi_\rho^I
    - \ft14 i \eps^{\mu\nu\rho}
        \, \bch A \psi_\mu^I
       \, \bch B \gamma_\nu \psi_\rho  \, \JAB    .
\label{N-Lgrav-inserted}
\eeql

The contribution to the kinetic term of the fermion is slightly
more complicated than in the $N=2$ case, because the covariant
derivative acts on $\J$, too:
\beql
   D^\neu_\mu ( \bch A \gamma^a \ch B \, \JAB ) &=&
   2 \bch A \gamma^a D_\mu^\neu \ch B \, \JAB   +
     \bch A \gamma^a \ch B \, D_\mu \JAB , \zeile
    D_\mu \JAB  &=&
   \ft14 Q_\mu^{IJ} \, \big[ \G^{IJ}, \J \big]_{\dot A \dot B}
      + Q_\mu^\a \, \big[ \G^\a , \J \big]_{\dot A \dot B}  .
\label{N-derivative-J}
\eeql
If the last expression does not vanish, i.e. if $\J$
does not commute with the whole gauge group $H$, there will be a
mixing of complexified spinors and their Dirac conjugates. Since
$\J$ generates a $\grp U(1)$ subgroup of $H$, $D_\m \JAB =0$ whenever
$H$ has a $\grp U(1)$ factor, i.e. whenever the target space is a
Kaehler manifold. This is the case for the theories with
$N=2 \mod 4$ \cite{dWNT}. For $N\neq 2 \mod 4$, manifest invariance
under those generators of $H$ which do not commute with $\J$ is lost.
For instance, the manifest $\grp SO(16)$ invariance of $N=16$
supergravity is thereby broken to $\grp SU(8)\times \grp U(1)$.

Inserting everything into the Lagrangian given in \cite{dWNT}
and dropping the primes on the redefined spin connection, we get
\beql
  \Lag0 & = &  \ft14 \eps^{\mu\nu\rho} \euo \mu a F_{\nu\rho a}
    +\ft12 \eps^{\mu\nu\rho} \bar \psi_\mu^I D_\nu \psi_\rho^I\,,
            \zeile
  \Lag1 & = & -\ft14 eg^{\mu\nu} \suco P_{A\mu} \suco P_{A\nu}
                + \ft12 \eps^{\mu\nu\rho}  \suco P_{A\mu}
   \Gam I{A \dot C} \, \bch C \gamma_\rho \psi^I_\nu\,
  \zeile &&{}\ \ \ \   + \ft14 \eps^{\mu\nu\rho}  \,
(\Gam I{A \dot B} \Gam J{A \dot C} + i \delta^{IJ} \J_{\dot B \dot C} )
      \, \bch B            \psi^I_\mu
      \, \bch C \gamma_\rho \psi^J_\nu ,\zeile
  \Lag2 & = & - \ft12 e \, \bch A \gamma^\mu D_\mu \ch B
        \, (\delta_{\dot A \dot B} + i\JAB  )
        - \ft14 \, \bch A \ch B \, \J_{\dot B \dot C}
                      \, \bch C \ch D \, \J_{\dot D \dot A}
  \zeile &&{} \ \ \ \
         - \ft14 i e \, \bch A \gamma^\mu \ch B
          \, D_\mu \JAB   +
       \mbox{ other $\chi^4$ terms }.
\eeql
where we have decomposed the Lagrangian as in \rf{2-Lagrange}.

The momenta of the dreibein, spin connection, gravitinos and
fermions are now
\beql
   \epuls ai & =&  \deltadelta \Lagrange /   \dot \euo ia / = 0,
           \zeile
   \Apuls ai &= &\deltadelta \Lagrange /   \dot \Ao ia / =
                     \ft12 \eps^{ij} e_{ja} ,
           \zeile
   \bar \pi^{iI} &=& \deltadelta \Lagrange /  \dot \psi_i^I / =
      - \ft12 \eps^{ij} \bar \psi_j^I ,
           \zeile
   \blam A &=& \deltadelta \Lagrange / \dot \chi^{\dot A}  / =
       \ft12   \bch B \, e \gamma^t \,
     (\delta_{\dot B \dot A}+ i\J_{\dot B \dot A}).
\label{N-second-class-momenta}
\eeql
Observe that $\blam A$ is complex and thus no longer a Majorana
spinor. The Poisson brackets read
\beqx
  \pois { \euo ia } { \epuls bj } = \delta^a_b \delta_i^j , \ \ \ \
  \pois { \Ao ia } { \Apuls bj } = \delta^a_b \delta_i^j ,
\eeqx
\beq
  \pois { \bar \l_{\dot A \a}  }{ \chi^{\dot B}_\b } =
       - \delta^{\dot B}_{\dot A} \delta_{\a \b}  , \ \ \ \
  \pois { \bar \pi^{iI}_\alpha }{ \psi^I_{j\beta} } =
       - \delta_j^i \delta^{IJ} \delta_{\alpha\beta}  ,
\eeq
and the full set of second class constraints is
\beql
       P_a^i  & :=&    \epuls ai,  \zeile
       Z_a^i  & :=&    \Apuls ai - \ft12 \eps^{ij} e_{ja} ,\zeile
   \bar\Lambda_{\dot A \a} & :=&  \bar \l_{\dot A \a} -
 \ft12 \, \bar \chi^{\dot B}_\b (C\, e\g^t)_{\b \a }
            \, ( \delta_{\dot B \dot A} + i \J_{\dot B \dot A} ),\zeile
   \bar\Gamma^{iI}_\alpha & := &  \bar\pi^{iI}_{\alpha}
            + \ft12 \eps^{ij} \, \psi_{j\beta}^I C_{\beta\alpha}.
\label{N-second-class-constraints}
\eeql
To obtain the Dirac brackets one has to invert the matrix of
Poisson brackets of these constraints.
The nonvanishing components of this matrix are
\beql
  \pois { P_a^i } { Z_b^j } &=& - \ft12 \eta_{ab} \eps^{ij},\zeile
  \pois { P_a^i } {\bar \Lambda_{\dot B \b} } &=&
  - \ft12 \, \chi^{\dot A}_\a   \,
           ( C  \g^c )_{\a \b} \, (\delta_{\dot A \dot B} + i \JAB )
     \, \eps_{abc} \eps^{ij} \euo jb ,\zeile
  \pois {\bar \Lambda_{\dot A \a} }{\bar \Lambda_{\dot B \b} } & =&
     (C\, e\g^t)_{\a \b} \, \delta_{\dot A \dot B} ,\zeile
  \pois {\bar\Gamma^{iI}_\alpha }{ \bar \Gamma^{jJ}_\beta }  &=&
      \eps^{ij} \delta^{IJ} C_{\alpha\beta} .
\label{N-Poisson-matrix}
\eeql
The inverse matrix is found to be
\beql
  C( P_a^i , Z_b^j ) &=&   2 \eps_{ij} \eta^{ab} , \zeile
  C( Z_a^i , Z_b^j ) &=&
     - \eps_{ij} \, h^{-1} \, e\eoo ta \, e \eoo tb  \,
      \bch A \ch B \,  (\delta_{\dot A \dot B}
          + \J_{\dot A \dot C} \J_{\dot C \dot B} ) ,\zeile
  C( Z_a^i , \bar \Lambda_{\dot A \a}) & = &
   h^{-1} \eps^{abc} \euu ib \, ( \delta_{\dot A \dot B} - i\JAB )
          ( e\g^t \, C)_{\a \b } \, \chi^{\dot B}_\b  , \zeile
  C( \bar \Lambda_{\dot A\a} , \bar\Lambda_{\dot B \b} ) & =&
   - \delta_{\dot A \dot B} \, h^{-1} \,( e\g^t\, C^{-1})_{\a \b} ,
           \zeile
  C( \bar \Gamma^{iI}_\alpha , \bar \Gamma^{jJ}_\beta ) & = &
       -\eps_{ij} \, \delta^{IJ} \, C_{\alpha\beta}^{-1} .
\label{N-inverse-Poisson-matrix}
\eeql
The bracket between $\Ao ia$ and $\Ao jb$ is easily seen to vanish
if $C(Z_a^i,Z_b^j)$ vanishes. This is the case if and only
if $\JAB \J_{\dot B \dot C} = - \delta_{\dot A \dot C}$. Hence
$\JAB$ is indeed a complex structure as previously asserted.
With $\J^2 = - \eins$, the combinations
\beq
{\cal P}^\pm_{\dot A \dot B} := \ft12 \left( \delta_{\dot A \dot B}
 \pm i \JAB \right)
\eeq
become projection operators acting on the fermions. This is important
because only then the number of physical fermionic degrees of freedom
stays the same after complexification: we are simply
trading $d$-dimensional real spinors for
$d\over 2$-dimensional complex ones. However,
as already pointed out above, the complexified spinors will no
longer transform linearly under the group $H$ unless $N=2 \mod 4$.

Using the formula \rf{2-Dirac-formula} we obtain
\beql
   \dirac { \Ao ia }{\Ao jb } = 0, \ \ & & \ \
   \dirac { \euo ia }{ \euo jb } = 0, \zeile
   \dirac { \Ao ia}{ \euo jb } & = & 2 \eps_{ij} \eta^{ab}.
\eeql
Furthermore,
\beq
   \dirac {\Ao ia } {\bar \l_{\dot A \a} } = 0 , \ \ \ \
   \dirac {\bar \l_{\dot A \a} }{ \bar \l_{\dot B \b} } = 0.
\eeq
This shows that the $\blam A$'s are indeed good canonical variables
as their brackets decouple from $\Ao i a$. On the other hand, the
original real spinors $\ch A$ are {\it not}, as they mix with each
other and the spin connection under Dirac brackets. However, the
correct variables are now easy to guess; they are
\beq
   \eta^{\dot A} :=
         {\cal P}^+_{\dot A \dot B}  \ch B = \ft12 \,
   ( \delta_{\dot A \dot B} + i \JAB  )\,  \ch B,
\label{N-eta-definition}
\eeq
and are related to $\blam A$ by $\blam A= \bar \eta^{\dot A}
e\g^t$, where the bar on $\eta$ denotes Dirac conjugation.
It is now straightforward to check that
\beq
   \dirac {\Ao ia  } {\eta^{\dot A}_\a  } = 0   , \ \ \ \
   \dirac { \eta^{\dot A}_\a  }{  \eta^{\dot B}_\b  } = 0 , \ \ \ \
   \dirac { \eta^{\dot A}_\a  }{ \bar \l_{\dot B \b }  }
     = - \delta^{\dot A}_{\dot B}  \delta_{\a \b} .
\eeq
This completes the proof that the canonical variables can be
decoupled by the redefinition \rf{N-new-connection}


\begin{thebibliography}{9}

\bibitem{WDW}
J.A. Wheeler, in {\sl Relativity, Groups and Topology},
eds. C. DeWitt and B. DeWitt, Gordon and Breach, New York, 1964; \\
B. DeWitt, Phys.Rev. {\bf 160} (1967) 113; {\bf 162} (1967) 1195

\bibitem{Isham}
C.J. Isham, in {\sl  Recent Aspects of Quantum Fields}, Proceedings,
Schladming 1991, eds. H. Mitter and H. Gausterer, Springer Verlag,
1991

\bibitem{Ashtbook}
A. Ashtekar,
{\sl Lectures on Non-Perturbative Canonical Gravity}
World Scientific, Singapore, 1991

\bibitem{JacSmol}
T. Jacobson and L. Smolin,
Nucl. Phys. {\bf B299} (1988) 295

\bibitem{RovSmol}
C. Rovelli and L. Smolin,
Nucl. Phys. {\bf B331} (1990) 80

\bibitem{Matschull}
H.J. Matschull, {\it Solutions to the Wheeler DeWitt Constraint of
       Canonical Gravity Coupled to Scalar Matter Fields},
       gr-qc/9305025

\bibitem{GSW}
M. Green, J.H. Schwarz and E. Witten, {\sl Superstring Theory I,II},
Cambridge University Press, 1987

\bibitem{GM}
P. Ginsparg and G. Moore, {\it Lectures on 2D Gravity and 2D String
  Theory}, preprint YCTP-P23-92, hep-th/9304011

\bibitem{Nic91}
H. Nicolai, Nucl. Phys. {\bf B353} (1991) 493

\bibitem{MN}
H. Nicolai and H.J. Matschull, {\it Aspects of Canonical Gravity and
            Supergravity}, preprint DESY 92-099, to appear in
            J. of Phys. and Geom.

\bibitem{Forger}
M. Forger, J. Laartz and U. Schaper,
      Comm. Math. Phys. {\bf 146} (1992) 397 and
      Freiburg Preprint THEP-92-24 (1992)

\bibitem{CJ}
 E. Cremmer, S. Ferrara and J. Scherk, Phys.Lett. {\bf B74} (1978) 61;\\
 E. Cremmer and B. Julia, Nucl.Phys. {\bf B159} (1979) 141

\bibitem{Geroch}
R. Geroch, J. Math. Phys. {\bf 12} (1971) 918, {\bf 13} (1972) 394

\bibitem{Schlad}
B. Julia, in {\sl Superspace and Supergravity}, eds. S.W. Hawking
and M. Rocek, Cambridge University Press, 1980;  \\
P. Breitenlohner and D. Maison, Ann. Inst. Poincar\'e {\bf 46}
(1987) 215; \\
H. Nicolai, in {\sl  Recent Aspects of Quantum Fields}, Proceedings,
Schladming 1991, eds. H. Mitter and H. Gausterer, Springer Verlag,
1991

\bibitem{DesJacHoo}
S. Deser, R. Jackiw and G. 't Hooft, Ann.Phys. {\bf 152} (1984) 220; \\
S. Deser and R. Jackiw, Ann.Phys. {\bf 153} (1984) 405; \\
J. Abbott, S. Giddings and K. Kuchar, Gen.Rel.Grav.{\bf 16} (1984) 751

\bibitem{Witten88}
E. Witten, Nucl.Phys. {\bf B311} (1988) 46

\bibitem{Martin}
S.P. Martin, Nucl. Phys. {\bf B327} (1989) 178; \\
S. Carlip, Phys. Rev. {\bf D42} (1990) 2647

\bibitem{Asht et al}
A. Ashtekar, V. Husain, C. Rovelli, J. Samuel and L. Smolin,
Class. Quantum Grav. {\bf 6} (1989) L185.

\bibitem{Bengtsson}
 I. Bengtsson, Phys.Lett. {\bf B220} (1989) 51

\bibitem{oldWitten}
K. Koehler, F.Mansouri, C. Vaz and L. Witten,
Nucl.Phys. {\bf B448} (1991) 373,
    {\bf B341} (1990) 167 and {\bf B358} (1991) 677

\bibitem{Teitelboim}
C. Teitelboim, Phys. Rev. Lett. {\bf 38} (1977) 1106

\bibitem{DKS}
S. Deser, J.H. Kay and K.S. Stelle, Phys.Rev. {\bf D16} (1977) 2448; \\
E.S. Fradkin and M.A. Vasiliev, Phys.Lett. {\bf B72} (1977) 70;  \\
M. Pilati, Nucl.Phys. {\bf B132} (1978) 138; \\
T. Jacobsen, Class. Quant. Grav. {\bf 5} (1988) 923

\bibitem{D'Eath1}
P. D. D'Eath, Phys. Rev. {\bf D29} (1984) 2199

\bibitem{Graham}
S. Elitzur, A. Forge and E. Rabinovici,
       Nucl.Phys. {\bf B274} (1986) 60; \\
P.D. D'Eath and D.I. Hughes, Phys. Lett. {\bf 214B} (1988) 498; \\
R. Graham, Phys. Rev. Lett. {\bf 67} (1991) 1381,
           Phys. Lett. {\bf B277} (1992) 393;     \\
P.D. D'Eath, S.W. Hawking and O. Obregon,
       Phys. Lett. {\bf B300} (1993) 44


\bibitem{Wess}
J. Bagger and J. Wess, {\sl Supersymmetry and Supergravity},
      Princeton University Press, 1983; \\
P.C. West, {\it Introduction to Supersymmetry and Supergravity},
      World Scientific, 1986

\bibitem{PvN}
 P. van Nieuwenhuizen, Phys.Rep. {\bf 68} (1981) 189

\bibitem{Brueg}
H. Kodama, Phys.Rev. {\bf D42} (1990) 2548; \\
B. Bruegmann, J. Pullin, Nucl.Phys. {\bf B363} (1991) 221; \\
B. Bruegmann, R. Gambini, J. Pullin,
Syracuse preprints SU-GP-92/1-1(1992) and SU-GP-92/3-1

\bibitem{dWNT}
B. de Wit, H. Nicolai and A. Tollsten, Nucl. Phys. {\bf B392} (1993) 3

\bibitem{Dirac}
P.A.M. Dirac,
{\sl Lectures on Quantum Mechanics},
Academic Press, New York, 1965

\bibitem{ADM}
R. Arnowitt, S. Deser and C.W. Misner,
%{\sl The Dynamics of General Relativity},
in {\sl Gravitation: An Introdoction to Current Research},
ed. L. Witten, Wiley, New York, 1962

\bibitem{MTW}
C.W. Misner, K.S. Thorne and J.A. Wheeler,
{\sl Gravitation},
Freeman, New York, 1973

\bibitem{HRT}
 A.J. Hanson, T. Regge and C. Teitelboim, {\sl Constrained
 Hamiltonian Systems}, Accademia Nazionale dei Lincei, Roma, 1976; \\
 M. Henneaux, Phys.Rep.{\bf 126} (1985) 1

\bibitem{DZ}
S. Ferrara, D.Z. Freedman and P. van Nieuwenhuizen, Phys.Rev. {\bf B13}
(1976) 3214; \\
S. Deser and B. Zumino, Phys.Lett.62B(1976)335

\bibitem{MSS}
N. Marcus, A. Sagnotti and J.H. Schwarz, Nucl. Phys. {\bf B243}
          (1984) 145

\bibitem{CH}
R. Courant and D. Hilbert, {\it Methoden der Mathematischen Physik},
    Springer Verlag, 1968


\bibitem{Sano}
T. Sano and J. Shiraishi, {\it The Nonperturbative Canonical
Quantization of the $N=1$ Supergravity}, Tokyo Preprint UT-622 (1992)

\bibitem{D'Eath2}
P. D'Eath, {\it Physical States in $N=1$ Supergravity},
   Cambridge preprint (1993)

\bibitem{Page}
D. Page, {\it Inconsistency of Canonically Quantized $N=1$
   Supergravity?}, Alberta preprint Thy-28-93, hep-th/9306032

\end{thebibliography}
\end{document}